\documentclass[lettersize,journal]{IEEEtran}
\usepackage{amsmath,amsfonts}
\usepackage{algorithmic}
\usepackage{algorithm}
\usepackage{array}
\usepackage{textcomp}
\usepackage{stfloats}
\usepackage{url}
\usepackage{verbatim}
\usepackage{graphicx}
\usepackage{cite}

\usepackage{booktabs}

\usepackage{color}
\usepackage{caption}
\usepackage{subcaption}
\usepackage{tabularx}
\usepackage[switch]{lineno}
\begin{document}

\title{Unrevealed Threats: A Comprehensive Study of the Adversarial Robustness of Underwater Image Enhancement Models}

\author{Siyu Zhai, Zhibo He, Xiaofeng Cong, Junming Hou, Jie Gui,~\IEEEmembership{Senior Member},~IEEE, Jian Wei You,~\IEEEmembership{Senior Member},~IEEE, Xin Gong,~\IEEEmembership{Member},~IEEE, James Tin-Yau Kwok,~\IEEEmembership{Fellow},~IEEE, Yuan Yan Tang,~\IEEEmembership{Life Fellow},~IEEE
        \thanks{
                 (Corresponding author: J. Gui.)

                 S. Zhai is with the School of Cyber Science and Engineering, Southeast University, Nanjing 210000, China (e-mail: 220224754@seu.edu.cn).

                 Z. He is with Information Management and Information Systems, Xian Jiaotong Liverpool University, Suzhou 215123, China (e-mail: Zhibo.He21@student.xjtlu.edu.cn).

                 X. Cong is with the School of Cyber Science and Engineering, Southeast University, Nanjing 210000, China (e-mail: cxf\_svip@163.com).

                J. Hou and J. You are with the State Key Laboratory of Millimeter Waves, School of Information Science and Engineering, Southeast University, Nanjing 210096, China (e-mails: junming\_hou@seu.edu.cn, jvyou@seu.edu.cn).

                J. Gui is with the School of Cyber Science and Engineering, Southeast University and with Purple Mountain Laboratories, Nanjing 210000, China (e-mail: guijie@seu.edu.cn).

                X. Gong is with the School of Cyber Science and Engineering, Southeast University, Nanjing 210096, China (e-mail: xingong@seu.edu.cn)

                J. Kwok is with the Department of Computer Science and Engineering, The Hong Kong University of Science and Technology, Hong Kong 999077, China. (e-mail: jamesk@cse.ust.hk).

                Y. Tang is with the Department of Computer and Information Science, University of Macau, Macau 999078, China (e-mail: yuanyant@gmail.com).
}
}

\markboth{Journal of \LaTeX\ Class Files,~Vol.~14, No.~8, August~2021}%
{Shell \MakeLowercase{\textit{et al.}}: A Sample Article Using IEEEtran.cls for IEEE Journals}

\makeatletter
\def\ps@IEEEtitlepagestyle{
  \def\@oddfoot{\mycopyrightnotice}
  \def\@evenfoot{}
}
\def\mycopyrightnotice{
  {\footnotesize
  \begin{minipage}{\textwidth}
  \centering
  Copyright~\copyright~20xx IEEE. Personal use of this material is permitted. \\ 
   However, permission to use this material for any other purposes must be obtained from the IEEE by sending a request to pubs-permissions@ieee.org.
  \end{minipage}
  }
}
\maketitle

\begin{abstract}
  Learning-based methods for underwater image enhancement (UWIE) have undergone extensive exploration.
  However, learning-based models are usually vulnerable to adversarial examples so as the UWIE models.
  To the best of our knowledge, there is no comprehensive study on the adversarial robustness of UWIE models, which indicates that UWIE models are potentially under the threat of adversarial attacks. 
  In this paper, we propose a general adversarial attack protocol. We make a first attempt to conduct adversarial attacks on five well-designed UWIE models on three common underwater image benchmark datasets.
  Considering the scattering and absorption of light in the underwater environment, there exists a strong correlation between color correction and underwater image enhancement. 
  On the basis of that, we also design two effective UWIE-oriented adversarial attack methods Pixel Attack and Color Shift Attack targeting different color spaces.
  The results show that five models exhibit varying degrees of vulnerability to adversarial attacks and well-designed small perturbations on degraded images are capable of preventing UWIE models from generating enhanced results. 
  Further, we conduct adversarial training on these models and successfully mitigated the effectiveness of adversarial attacks. 
  In summary, we reveal the adversarial vulnerability of UWIE models and propose a new evaluation dimension of UWIE models. 
\end{abstract}

\begin{IEEEkeywords}
  Underwater Image Enhancement, Adversarial Robustness, Attack, Defense
\end{IEEEkeywords}

\section{Introduction}
\begin{figure}[t]
    \begin{subfigure}[t]{0.45\textwidth}
        \flushleft
       \includegraphics[width = \textwidth]{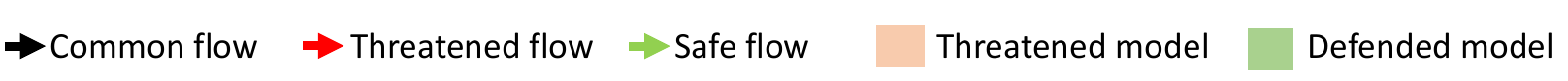}
    \end{subfigure}
    \begin{subfigure}[t]{0.235\textwidth}
        \flushleft
       \includegraphics[width = \textwidth]{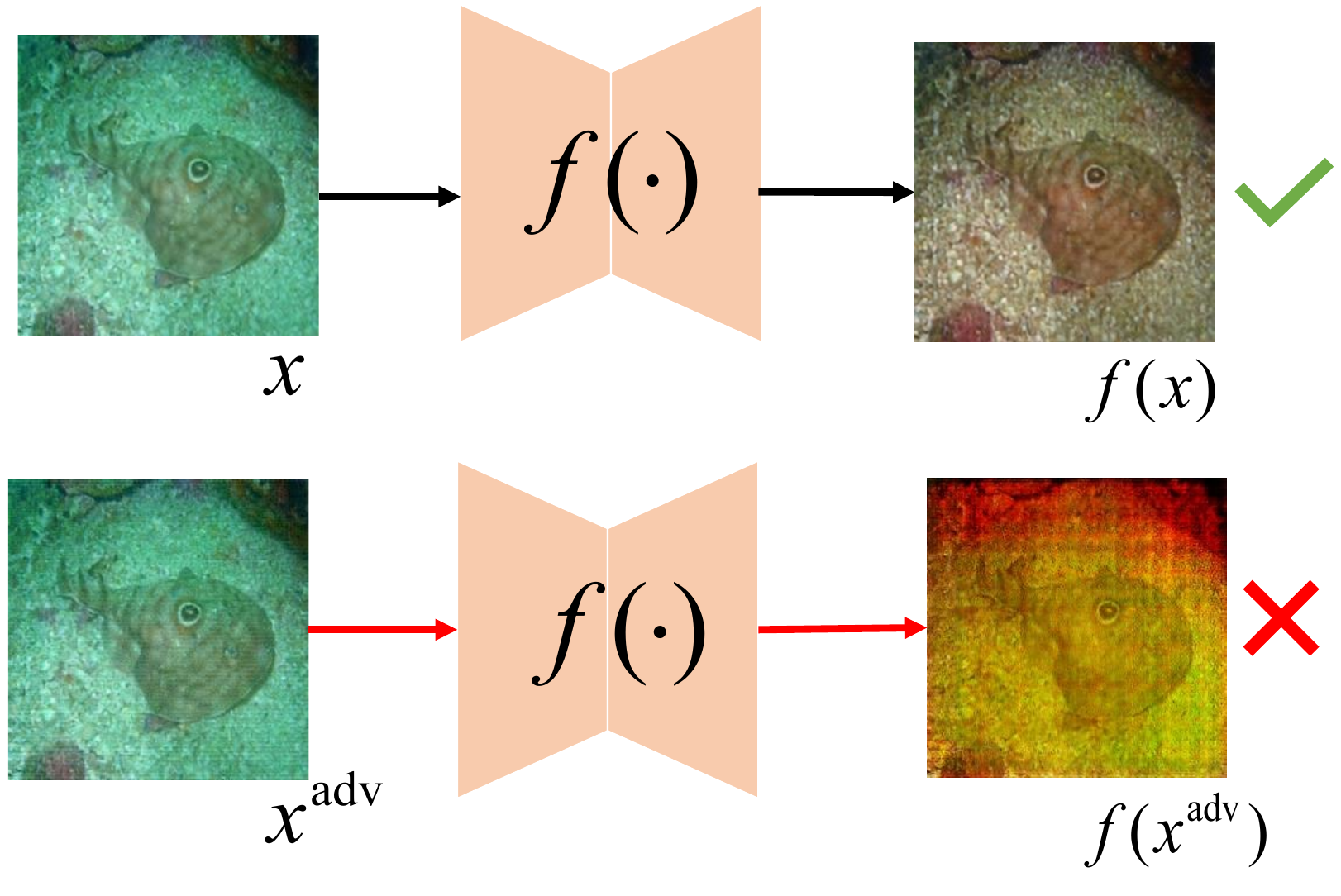}
       \caption{Without defense.}
    \end{subfigure}
    \begin{subfigure}[t]{0.235\textwidth}
        \flushleft
       \includegraphics[width = \textwidth]{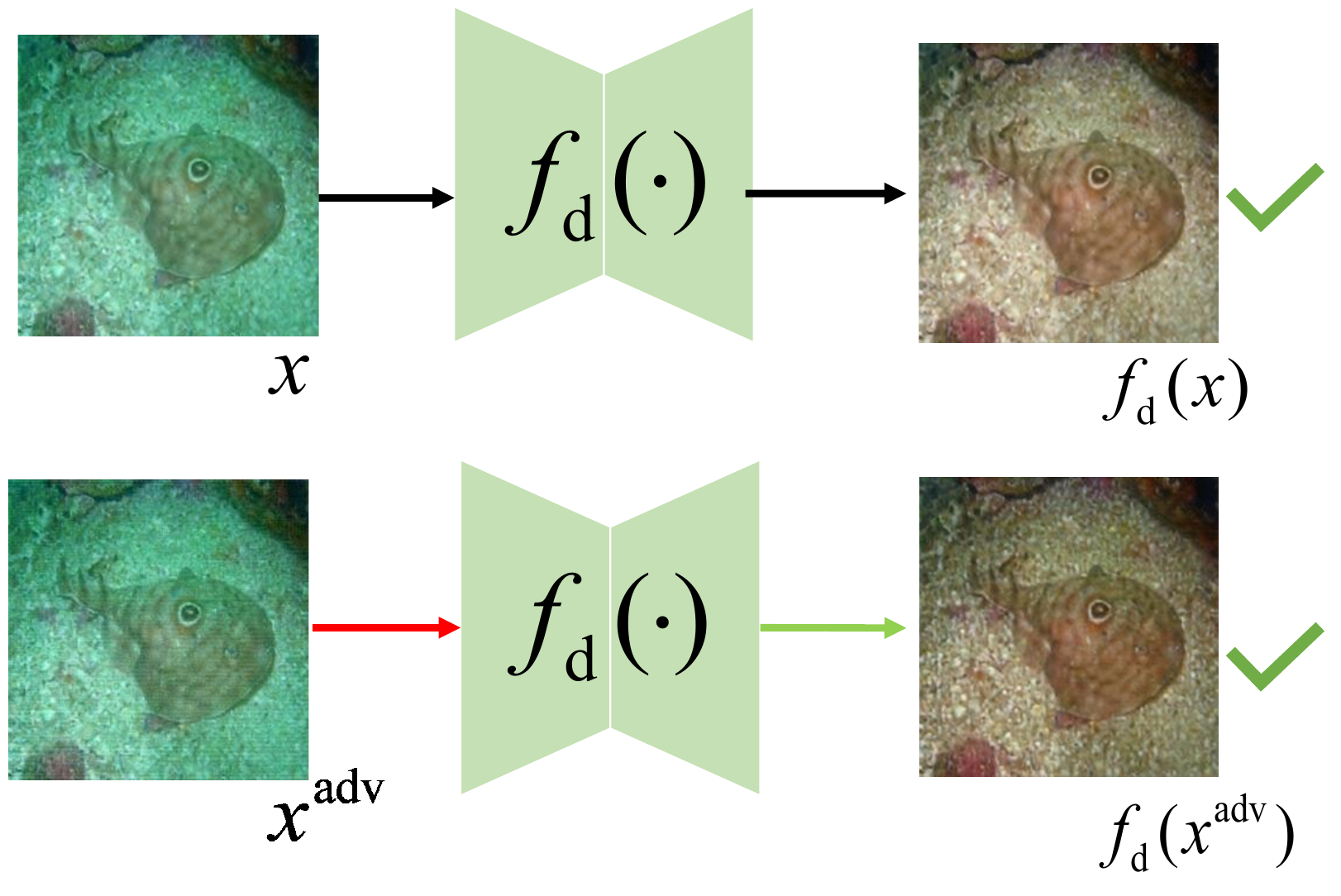}
       \caption{With defense.}
    \end{subfigure}
    \\
    
    \begin{subfigure}[t]{0.235\textwidth}
        \flushright
       \includegraphics[width = \textwidth]{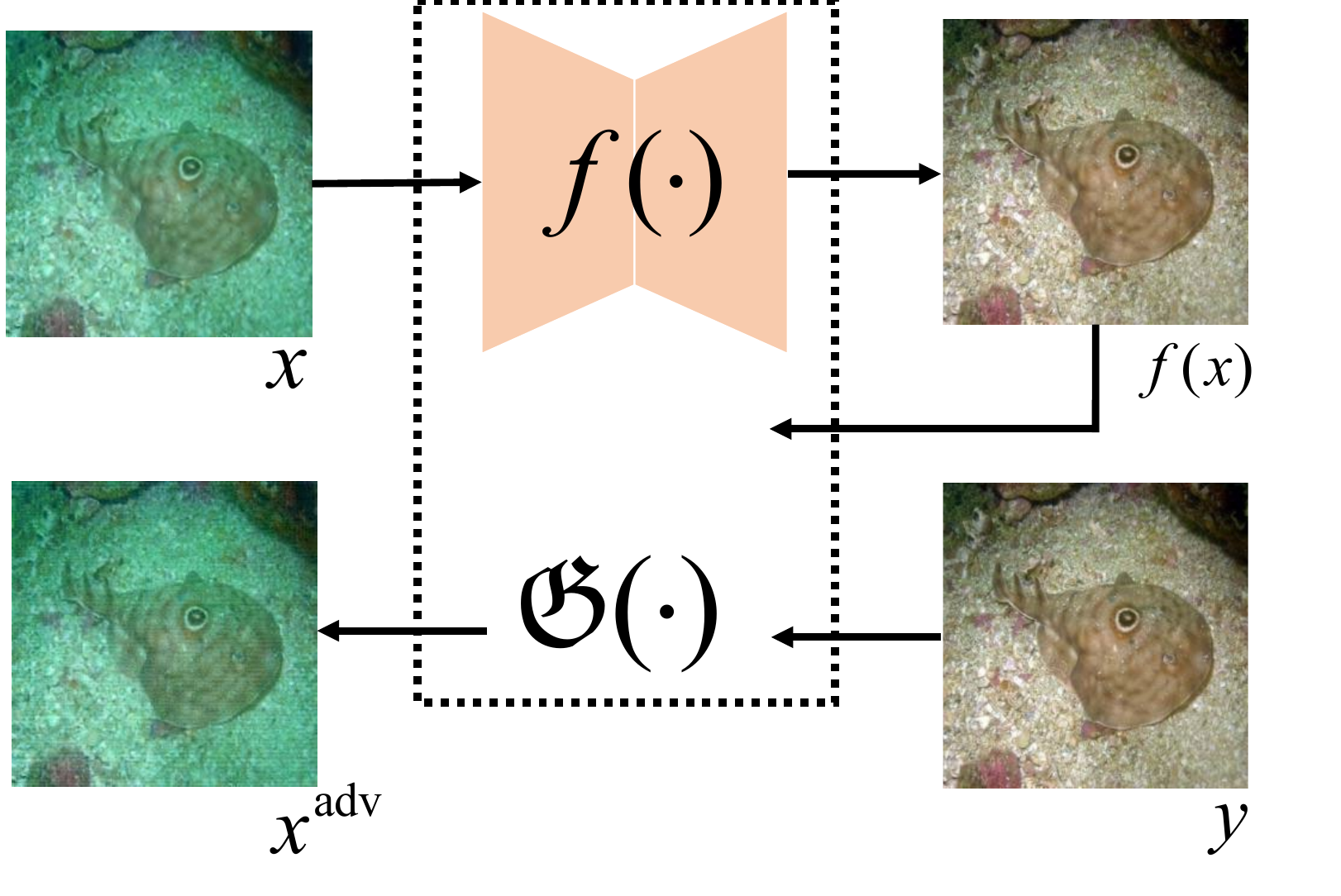}		
       \caption{Attack process.}
    \end{subfigure}
    \begin{subfigure}[t]{0.235\textwidth}
        \flushright
       \includegraphics[width = \textwidth]{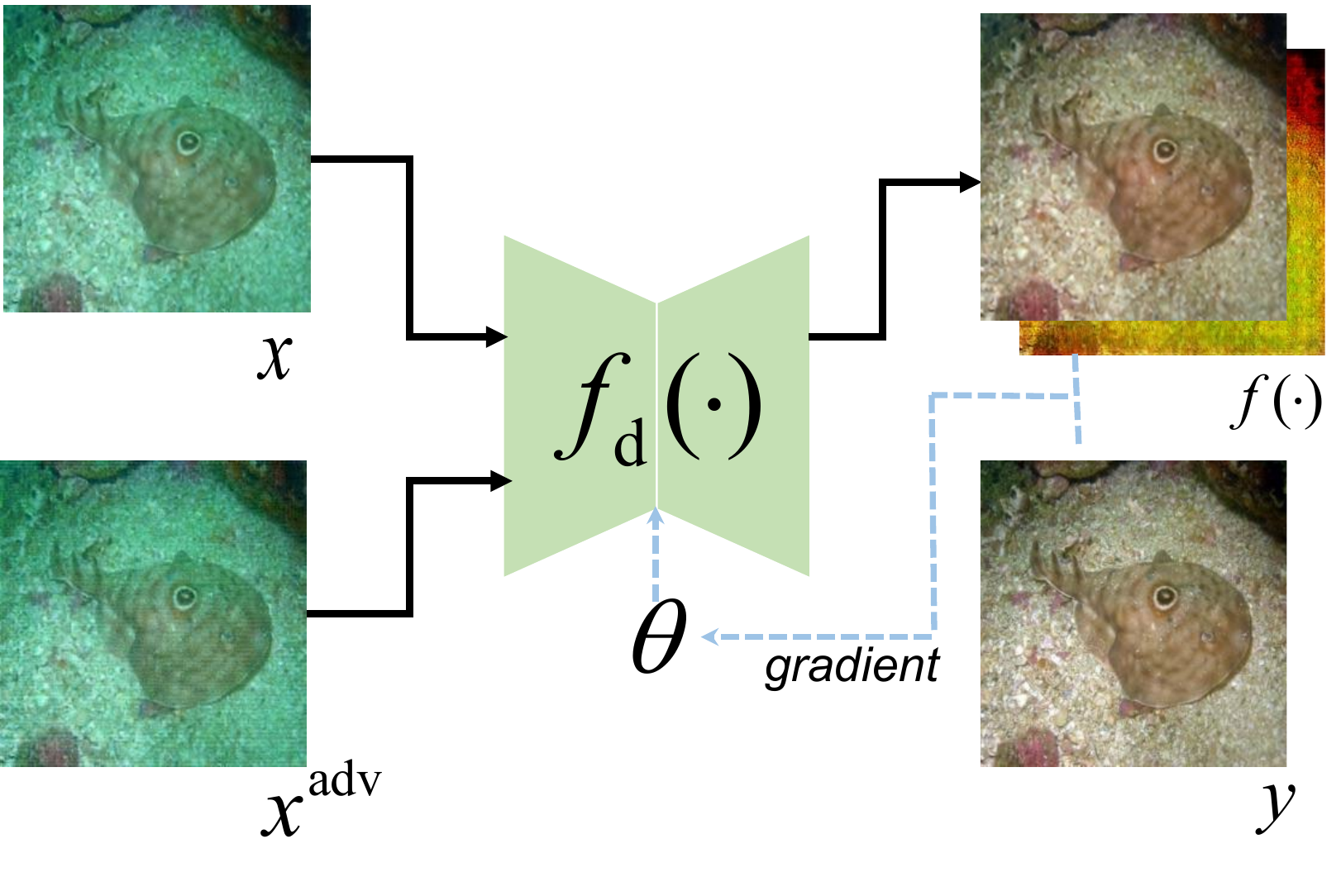}		
       \caption{Defense process.}
    \end{subfigure}
    \caption{\label{fig:over} Overall pipline of our study. In (a), the adversarial example $x^{\text{adv}}$makes the model $f(\cdot)$ generate unacceptable results. With our defense strategy, the defended model $f_{\text{d}}(\cdot)$ resists such attacks in (b).
    (c) demonstrates how we generate the adversarial example by utilizing the ground-truth $y$, where $\mathfrak{G} (\cdot)$ represents the adversarial attack method. 
    In the defense process in (d), $x$ and $x^{\text{adv}}$ are both used for updating network parameters $\theta$ to help $f_{\text{d}}(\cdot)$ resist adversarial attacks.}
    \vspace{-2\baselineskip}
\end{figure}

Underwater environments are of great importance 
for water covers 71\% of the earth's surface 
and provides natural habitats for most living organisms~\cite{mclellan2015sustainability}.
With abundant resources in the ocean, exploring, developing, and protecting the oceans have become active
research topics~\cite{alsakar2023underwater}. 
Due to the complex underwater environment and lighting conditions (wavelength-dependent
absorption, reflection, scattering),
underwater images often exhibit noticeable degradation. 
 These degraded underwater images often suffer from color distortion, low contrast, and blurriness~\cite{lin2022global,xue2023investigating,li2019underwater,jiang2022target,liu2022twin}.

Owing to the distinctive and intricate nature of the underwater imaging process, conventional enhancement methods tailored for alternative degradation scenarios,
such as low-light scenes~\cite{guo2020zero,li2021learning,jiang2022unsupervised} or foggy scenes~\cite{ijcai2022p160,wang2022adaptive} 
exhibit limited generalizability when applied to Underwater Image Enhancement (UWIE) task~\cite{zhou2023underwater,alsakar2023underwater,zhou2024hclr}.
Traditional UWIE methods~\cite{ghani2017automatic,sankpal2019underwater,priyadharsini2018wavelet,wang2022periodic}
have inferior and limited performance in many situations because of their hand-designed features or imperfect modeling.

With deep learning development in underwater image restoration,
 researchers have seen a major shift from traditional models to learning-based models~\cite{li2019underwater,yan2022attention,islam2020fast,chen2021underwater,9747758,wei2022uhd,10155564,zhou2023underwater}.
Though learning-based UWIE models have undergone extensive exploration, 
their adversarial robustness seems to be neglected. 
Since Goodfellow et al.~\cite{goodfellow2014explaining} pointed out that deep neural networks show their vulnerability,
studies have been devoted 
to the exploration of adversarial attacks and defenses~\cite{madry2018towards,wei2023cfa,croce2021mind}.

Recently, a few researchers have shown that 
restoration models on some low-level computer vision tasks are also subjected to the potential threat of adversarial examples,
varying in the field of deraining ~\cite{yu2022towards} and dehazing~\cite{gui2023adversarial}. 
Their research reveals the characteristics of adversarial attacks and defenses on specific low-level computer vision tasks.

However, to the best of our knowledge, there is no comprehensive study on the adversarial robustness of UWIE models. 
The complex underwater environment makes UWIE tasks distinct from other low-level computer vision tasks. 
This uniqueness also extends to the adversarial attacks and defenses of underwater image enhancement models.
We conducted a study on adversarial attacks and defenses for underwater image enhancement based on \textit{color space} information of underwater images.
We proposed two UWIE-oriented adversarial attacks Pixel Attack and Color Shift Attack based on RGB color space and YUV color space respectively.
The overall framework is demonstrated in Fig.~\ref{fig:over}. We perform adversarial attacks 
on five well-designed UWIE models and evaluate their adversarial robustness on three common underwater image datasets. 
Our main contributions are summarized below.
\begin{itemize}
  \item We first examined the adversarial robustness of learning-based UWIE models and 
        studied the common process of adversarial robustness analysis on UWIE models, including attack, defense, and evaluation.
  \item We proposed two adversarial attack methods tailored for UWIE models and conducted adversarial attacks on five well-designed UWIE models. 
  \item We explored defense methods against adversarial attacks to UWIE methods by performing adversarial training. 
  The results showed its effectiveness in resisting adversarial attacks.

\end{itemize}

\section{Related Works}
\subsection{Learning-based Underwater Image Enhancement}
Underwater image enhancement (UWIE) is a typical low-level computer vision task~\cite{zhang2023hierarchical,zhuhierarchical,sharma2023wavelength,zhang2023rex,zhang2023waterflow},
 whose purpose is to generate high-quality images from degraded underwater images. 
Recently, learning-based UWIE models have drawn more attention for the amazing effectiveness of enhancement results
~\cite{yan2022attention,islam2020fast,chen2021underwater,9747758,wei2022uhd,li2019underwater}.
Li et al.~\cite{li2019underwater} were the first to conduct a comprehensive study
and analysis of underwater image enhancement with large-scale real-world images UIEB.
Islam et al.~\cite{islam2020fast} presented a large-scale dataset and guided UWIE tasks with several underwater-specific global information.
Chen et al.~\cite{chen2021underwater} combined the image formation model with deep learning to enhance the real-time inference capability of UWIE models, yet achieved better metrics.
Fu et al.~\cite{9747758} proposed water-type irrelevant encodings and a novel normalization scheme to free UWIE models from certain distortions. 
Yan et al.~\cite{yan2022attention} proposed an attention-guided dynamic multibranch neural network to obtain high-quality underwater images.
Wei et at.~\cite{wei2022uhd} utilized ultra-high definition (UHD) underwater imaging to rebuild informative color features of underwater images.
Zhou et al.~\cite{zhou2022underwater,zhou2023underwater} combined depth map estimation and backscatter elimination with unsupervised techniques and showed appealing results
in complicated underwater environments.
Though these researches have pushed the boundary of learning-based UWIE methods and made more delightful enhanced underwater images,
the adversarial robustness of UWIE models has not been comprehensively discussed.
\subsection{Adversarial Attacks and Defenses}

Though Deep Neural Networks (DNNs) have achieved remarkable success in many areas, 
their vulnerability against adversarial examples severely affects the application of deep learning models in safety-critical domains.
Underwater scenarios are among these domains. 
Goodfellow et al.~\cite{goodfellow2014explaining} found that DNNs can be easily fooled by perturbations that are unperceivable to humans.
Papernot et al.~\cite{papernot2017practical} summarized that if attackers fully acquire prior knowledge about the target model,
 including its structure, parameters, and training data, such attacks are named white-box attacks. 
 If attackers have no knowledge of the model structure or training set, such attacks are named black-box attacks.
Szegedy et al.~\cite{szegedy2013intriguing} proposed the Fast Gradient Sign Method (FGSM), a simple but effective method to generate adversarial examples.
FGSM establishes the fundamental form of generating adversarial examples based on a first-order gradient algorithm.
Madry~\cite{madry2018towards} presented a multi-step version of FGSM and named it Projected Gradient Descent (PGD). 
PGD can effectively find adversarial examples with perturbations restricted within an infinity norm ball.
Subsequent researchers have proposed a series of efficient methods for generating adversarial, such as Auto-Attack~\cite{croce2021mind}, DeepFool~\cite{moosavi2016deepfool} and C\&W~\cite{carlini2017towards}.

To resist the threat from adversarial examples, several defense approaches have been proposed~\cite{bai2019hilbert,das2017keeping,mo2022adversarial,papernot2016distillation}, 
and adversarial training is the most popular among them. Adversarial training generates adversarial examples during the training phase and
the generated adversarial examples will be incorporated into the model training. 

\subsection{Adversarial Robustness and Restoration Models}

Discussions about adversarial attacks and defenses are focused on DNNs on high-level tasks.
Initially, researchers primarily employed restoration models as a method for attacking or defending high-level task models.
Mustafa et al.~\cite{mustafa2019image} proposed a novel approach to apply super-resolution networks to destroy adversarial patterns.
Some researchers~\cite{zhai2020s,gao2021advhaze} attempted to generate specific rain streaks or haze as adversarial perturbations to attack DNN models.
Recently, researchers have noted research on the adversarial robustness of generative networks tailored for low-level tasks. 
These low-level tasks differ from high-level tasks in their model outputs and objectives. 
Yu et al.~\cite{yu2022towards} comprehensively illustrated the robustness of rain removal models. 
They proposed several attacks specific to rain removal and systematically analyzed the robustness of model components.
Song et al.~\cite{song2023robust} focused on single-image reflection removal (SIRR) and built an SIRR model that narrows the gap between original examples and adversarial examples.
Gui et al.~\cite{gui2023adversarial} designed several attack methods on dehazing tasks and defined a new challenging problem for image dehazing.

However, the adversarial robustness of UWIE models has not been comprehensively researched. 
In this paper, we focus on white-box attacks on UWIE models and conduct adversarial training on well-designed UWIE models and commonly used underwater image datasets to
explore the adversarial robustness of UWIE models.

\section{Adversarial robustness of UWIE methods}
\label{sec:methods}
\begin{table*}[]
    \centering
    \setlength{\tabcolsep}{0.32cm}
    \caption{PSNR and SSIM of undefended UWIE models for original examples and adversarial examples.  Bold results represent the lowest metric under adversarial attacks on this dataset.}
    \label{tab:1}
    \begin{tabular*}{\linewidth}{@{}ccccccccccccc@{}}
    \toprule
     &
      \multicolumn{4}{c}{UIEB} &
      \multicolumn{4}{c}{EUVP-S} &
      \multicolumn{4}{c}{EUVP-I} \\ \midrule
     &
      \multicolumn{2}{c}{PSNR$\uparrow$} &
      \multicolumn{2}{c}{SSIM$\uparrow$} &
      \multicolumn{2}{c}{PSNR$\uparrow$} &
      \multicolumn{2}{c}{SSIM$\uparrow$} &
      \multicolumn{2}{c}{PSNR$\uparrow$} &
      \multicolumn{2}{c}{SSIM$\uparrow$} \\ \midrule
    Metrics &
      clean &
      adv &
      clean &
      adv &
      clean &
      adv &
      clean &
      adv &
      clean &
      adv &
      clean &
      adv \\ \midrule
    \multicolumn{1}{c|}{ADMNNet} &
      \multicolumn{1}{l}{21.089} &
      \multicolumn{1}{l|}{14.349} &
      \multicolumn{1}{l}{0.890} &
      \multicolumn{1}{l|}{0.576} &
      26.710 &
      \multicolumn{1}{c|}{15.912} &
      0.883 &
      \multicolumn{1}{c|}{0.488} &
      24.438 &
      \multicolumn{1}{c|}{\textbf{12.990}} &
      0.841 &
      0.582\\
    \multicolumn{1}{c|}{FUnIEGAN} &
    19.087 &
      \multicolumn{1}{c|}{\textbf{11.296}} &
      0.817 &
      \multicolumn{1}{c|}{\textbf{0.508}} &
      27.489 &
      \multicolumn{1}{c|}{\textbf{6.599}}&
      0.865 &
      \multicolumn{1}{c|}{\textbf{0.315}} &
      23.161 &
      \multicolumn{1}{c|}{20.142} &
      0.806 &
      0.686 \\
    \multicolumn{1}{c|}{PhysicalNN} &
      18.568 &
      \multicolumn{1}{c|}{16.354} &
      0.804 &
      \multicolumn{1}{c|}{0.581} &
      25.912 &
      \multicolumn{1}{c|}{21.475} &
      0.848 &
      \multicolumn{1}{c|}{0.552} &
      23.571 &
      \multicolumn{1}{c|}{19.796} &
      0.808 &
      0.569 \\
    \multicolumn{1}{c|}{UHD} &
    17.658 &
      \multicolumn{1}{c|}{14.354} &
      0.783 &
      \multicolumn{1}{c|}{0.538} &
      22.714 &
      \multicolumn{1}{c|}{15.691} &
      0.815 &
      \multicolumn{1}{c|}{0.597} &
      20.294 &
      \multicolumn{1}{c|}{18.102} &
      0.745 &
      \textbf{0.501} \\
    \multicolumn{1}{c|}{SCNet} &
      21.552 &
      \multicolumn{1}{c|}{18.313} &
      0.900 &
      \multicolumn{1}{c|}{0.617} &
      27.820 &
      \multicolumn{1}{c|}{16.535} &
      0.878 &
      \multicolumn{1}{c|}{0.459} &
      24.562 &
      \multicolumn{1}{c|}{17.374} &
      0.826 &
      0.584 \\
     \bottomrule
    \end{tabular*}
    \end{table*}

To make our discussions more concise, we first make some agreements on essential concepts.
\begin{itemize}
    \item Original example: denoted as $x$, the original degraded underwater images from the dataset. They do not undergo any adversarial attack perturbations.
    \item Original output: denoted as $f(x)$, the enhanced result for original examples. $f(\cdot)$ represents the learning-based UWIE model, and it is parameterized with $\theta$.
    \item Adversarial attack: denoted as $\mathfrak{G} (\cdot)$, the method that adds imperceptible perturbations to original examples to make UWIE models generate degraded results.
    \item Adversarial perturbation: denoted as $\delta$, the imperceptible perturbations generated by adversarial attacks.
    \item Adversarial example: denoted as $x^{\text{adv}}$, the degraded underwater image to fool UWIE models.
    \item Defended model: denoted as $f_\text{d}(\cdot)$, UWIE models that undergo adversarial training.
    \item Ground truth: denoted as $y$, the label in the dataset.
\end{itemize}
The pipeline of our attacks and defenses is shown in Fig.~\ref{fig:over}. In attacking phase (Fig.~\ref{fig:over} (a)), adversarial examples generated by adversarial attacks in Fig.~\ref{fig:over} (c) are fed into UWIE models to generate degraded results.
With our defending tragedy in Fig.~\ref{fig:over} (d), robust UWIE models are able to resist adversarial attacks in Fig.~\ref{fig:over} (b).
\subsection{An Attack Protocol of UWIE Models}
In order to achieve the goal of attacking learning-based models~\cite{madry2018towards}, we add perturbations on the basis of maximizing the loss value.
Supposing that there is a UWIE model $f(\cdot)$ with network parameters ${\theta}$, we fomulate white-box attack for $f(\cdot)$ as a constrained optimization problem,
\begin{equation}
    \mathop{\max}\limits_{x^{\text{adv}}} \mathcal{L}(f(x^{\text{adv}}),y) \quad s.t. ||x^{\text{adv}}-x||_{\infty} < \epsilon,
    \label{eq:1}
\end{equation}
where $x \in \mathbb{R}^{3 \times W \times H}$ denotes original examples from dataset $\mathcal{D} $, and $\mathcal{L}$ is the loss function used by particular attack methods.
The constraint condition restricts  the adversarial preparation $\delta = x^{\text{adv}}-x$ to the infinity norm ball $\mathcal{B}_\infty(x,\epsilon)$. $x$ is the center of the sphere and $\epsilon$  is the radius.
This guarantees the adversarial example $x^{\text{adv}}$ is imperceptible to humans.
Eq.~\eqref{eq:1} can be rephrased with $\delta$,
\begin{equation}
    \mathop{\max}\limits_{\delta} \mathcal{L}(f(x+\delta),y) \quad s.t. ||\delta||_{\infty} < \epsilon.
    \label{eq:2}
\end{equation}

To solve the constrained maximization problem, we utilized a multi-step iteration method,
\begin{equation}
    x^{0} = x+ U(-\epsilon,\epsilon),
    \label{eq:6}
\end{equation}
\begin{equation}
    x^{t+1} = \Pi_{\mathcal{B}_\infty(x,\epsilon)}(x^{t}+\alpha \text{sgn}(\nabla_{x^{t}} \mathcal{L}(f(x^{t}),y))),
    \label{eq:7}
\end{equation}
\begin{algorithm}[tb]
    \caption{Our attack method, denoted as $\mathfrak{G} (\cdot)$}
    \label{alg:alg1}
    \textbf{Input}: A UWIE model $f(\cdot)$ parameterized by $\theta$, 
    a batch of degraded underwater images $x=\{x_{1},x_{2},\cdots,x_{m}\}$ with batch size $m$ 
    and corresponding ground truth $y=\{y_{1},y_{2},\cdots,y_{m}\}$\\
    \textbf{Parameter}: Attack parameters $\alpha,\epsilon$, max iteration number $T$\\
    \textbf{Output}: $x^{adv} = x^{T}$
    \begin{algorithmic}[1] 
        \STATE Let $t=0$.
        \STATE Let $x^{0}=x+U(-\epsilon,\epsilon)$     // initialize perturbations with uniform distribution
        \FOR{$t=0,1,\cdots,T-1$}
        \STATE $\delta = \alpha \cdot \text{sgn}(\nabla_{x^{t}} \mathcal{L}(f(x^{t}),y))$
        \STATE $\delta = \text{clip}(\delta,-\epsilon,\epsilon)$ // projection for infinity norm
        \STATE $x^{t+1} = x^{t}+\delta$
        \STATE $x^{t+1} = \text{clip}(x^{t+1},0,1)$ // make sure that pixels are valid
        \ENDFOR
        \STATE \textbf{return} 
        \label{alg:1}
    \end{algorithmic}
\end{algorithm}
\begin{algorithm}[tb]
    \caption{Adversarial training for a mini-batch}
    \label{alg:alg2}
    \textbf{Input}: A UWIE model $f(\cdot)$ parameterized by $\theta$, 
    a batch of degraded underwater images $x=\{x_{1},x_{2},\cdots,x_{m}\}$ with batch-size $m$ 
    and corresponding ground truth $y=\{y_{1},y_{2},\cdots,y_{m}\}$\\
    \textbf{Parameter}: model loss $\mathcal{L}_\text{model}$, Attack parameters $\alpha,\epsilon$, max iteration number $T$
    \begin{algorithmic}[1] 
        \STATE $x^{\text{adv}} = \mathfrak{G}(x)$
        \STATE Forward propagation for $f(x^{\text{adv}})$
        \STATE Usual forward propagation for $f(x)$
        \STATE $\mathcal{L} =\mathcal{L}_{{\text{model}}}+\lambda \mathcal{L}_{{\text{adv}}}$
        \STATE Update network parameters $\theta$ with $\mathcal{L}$ and the optimizer\STATE \textbf{return}
        \label{alg:2}
    \end{algorithmic}
\end{algorithm}
where $U(-\epsilon,\epsilon)$ is a uniform distribution on the interval $[-\epsilon,\epsilon]$, sgn$(\cdot)$ is the symbol function that returns the sign of each component,
$\Pi_{\mathcal{B}_\infty(x,\epsilon)} $ is the projection function that projects the tensor to the infinity norm ball $\mathcal{B}_\infty(x,\epsilon)$. 
The detailed procedure (denoted as $\mathfrak{G(\cdot)}$) is shown in Algorithm~\ref{alg:alg1}, where $ \text{clip}(x,a,b)$ clamps all elements in $x$ into interval $[a,b]$.

We then discuss the evaluation of the robustness of UWIE models. Generally, we use IQAs of adversarial outputs to assess the robustness of UWIE network $f(\cdot)$
is denoted as $R(\mathcal{D},f,\mathcal{S})$, 
\begin{equation}
    R(\mathcal{D},f,\mathcal{S}) = \mathbb{E}_{x\sim \mathbb{P}_{\mathcal{D}} }\mathcal{I} (f(x+\delta),y),
    \label{eq:3}
\end{equation}
where $\mathbb{P}_\mathcal{D}$ is the distribution of dataset $ \mathcal{D}$, 
$\mathcal{S}$ represents all settings of the adversarial method and $\mathcal{I}$ represents Image Quality Assessments (IQAs)~\cite{alsakar2023underwater} for UWIE models.

 We choose peak Signal-to-Noise Ratio (PSNR)~\cite{zhang2024robust} and structural similarity (SSIM)~\cite{wang2022semantic,zhou2023ugif} as our evaluation metrics $\mathcal{I}$.
In this paper, we calculate $\mathcal{I}$ over test examples to estimate $R(\mathcal{D},f,\mathcal{S})$ since accurate probability distribution is impossible to be obtained.
\subsection{Adversarial Attacks for UWIE Models}
    With the white-box attack protocol of UWIE models, we propose three adversarial attack methods that specifically target UWIE models, named Pixel Attack and Color Shift Attack.
    We will describe these three methods on the basis of the equations in Sec 3.1.
\paragraph{\bf Pixel Attack.}
Pixel Attack is designed to attack UWIE models by perturbating pixel values of the original examples.
To enlarge the pixel gap between original outputs and adversarial outputs, we maximize the $l_2$ distance between the adversarial output and the ground truth.
Pixel Attack is formulated as follows,
\begin{equation}
    \mathcal{L}(x^t,y) = \frac{1}{3 W  H}\sum_c||f(x^t)-y||_2.
    \label{eq:8}
\end{equation}
Pixel Attack can be treated as conducting adversarial attacks on RGB color space. 
Pixel Attack is a simple but effective adversarial attack for UWIE models, which significantly reduces the PSNR and SSIM of adversarial outputs.

\paragraph{\bf Color Shift Attack.}
Color Shift Attack is designed to attack UWIE models by shifting the color of the original examples. 
To extract color information from grayscale, we first transform the image from RGB space to YUV space~\cite{spl}. 
In the YUV color space, the Y component carries the grayscale version of the image, while the U and V components contain the color information.
We perform Color Shift Attack as follows,
\begin{equation}
    \mathcal{L}(x^t,y) = \frac{1}{2 W  H}\sum_{U,V}||YUV(f(x^t))-YUV(y)||_2,
    \label{eq:11}
\end{equation}

which maximizes the $l_2$ distance between the adversarial output and the ground truth in the YUV color space and 
only focus on the U and V components to make the adversarial perturbations pay more attention to the color information.

\subsection{Adversarial Defenses of UWIE Methods}

To defend learning-based UWIE models from adversarial examples, we attempt to use original examples and adversarial examples for training.
Adversarial examples are almost visually indistinguishable from the original, so they share the same ground truth. 
This helps reduce the distance between adversarial examples and original ones in latent space \cite{madry2018towards,zhang2019theoretically}.
We express adversarial training as a saddle point optimization problem,
\begin{equation}
    \mathop{\min\limits_{\theta}\max\limits_{\delta \in \mathcal{B}_\infty(\epsilon)}} E_{(x,y)\sim \mathbb{P}_\mathcal{D}}\mathcal{L}(f(x+\delta),y),
    \label{eq:4}
\end{equation}
where $\theta$ is the network parameters. 

In the training phase, we follow this paradigm in our paper,
\begin{equation}
    \mathcal{L}=\mathcal{L}_\text{model}+\lambda\mathcal{L}_{\text{adv}},
    \label{eq:5}
\end{equation}
where hyperparameter $\lambda$ is the coefficient of the adversarial regularization term. We divide the loss function into two parts. 
$\mathcal{L}_\text{model}$ is regular loss function for models without adversarial training, 
while $\mathcal{L}_{\text{adv}}$ is adversarial regularization for robustness. Eq.~\eqref{eq:5} indicates that we use both
original and adversarial examples in a training step, which helps achieve the balance between image quality and robustness~\cite{zhang2019theoretically}. 
We define our adversarial term as follows,

\begin{equation}
    \mathcal{L}_{\text{adv}} = \sum_{x}||f(x)-f(x^{\text{adv}})||_2,
    \label{eq:tra}
\end{equation}
$\mathcal{L}_{\text{adv}}$ makes the network generate similar outputs for original examples and adversarial examples, which means adversarial examples
have fewer chances to induce UWIE models to generate degraded images and thus achieve the effect of defense.

\section{Experiments and Results}
\label{sec:experiments}

\subsection{Basic Setup}

\paragraph{\bf Datasets.}
We perform our experiments on widely used UWIE datasets, including UIEB~\cite{li2019underwater}, EUVP-I, and EUVP-S~\cite{islam2020fast,cong2024underwater}.
Dataset UIEB contains 950 real underwater images (890 images of them with ground truth selected from several classic UWIE methods). 
We take 800 pairs of high-resolution images as our training set and the rest 90 pairs of size $256 \times 256$ are for testing.
EUVP-I and EUVP-S are subsets in the EUVP dataset with high quality. We split them into training sets including 3300 and 1976 images of size $256 \times 256$ respectively.
The test sets involve 400 and 218 images respectively.

\paragraph{\bf Benchmark models.} 
We perform adversarial attacks and defenses on five well-designed UWIE models including ADMNNet~\cite{yan2022attention}, FUnIEGAN~\cite{islam2020fast},
 PhysicalNN~\cite{chen2021underwater}, UHD~\cite{wei2022uhd} and SCNet~\cite{9747758}.
We train all pre-trained models of five models with original examples for 100 epochs and fix batch size to 6. We implement our experiments using the Pytorch with a single NVIDIA GeForce RTX 4090 GPU.

\begin{figure*}[t]
 
    \centering
    \includegraphics[width=9cm,height=0.5cm]{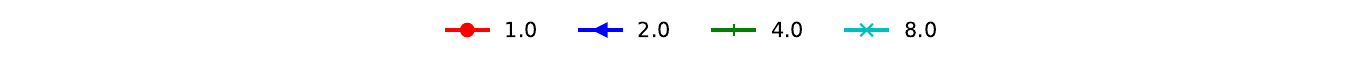}\\
    \begin{subfigure}[t]{0.19\textwidth}
        \centering
        \includegraphics[width = \textwidth, height=1.2\textwidth]{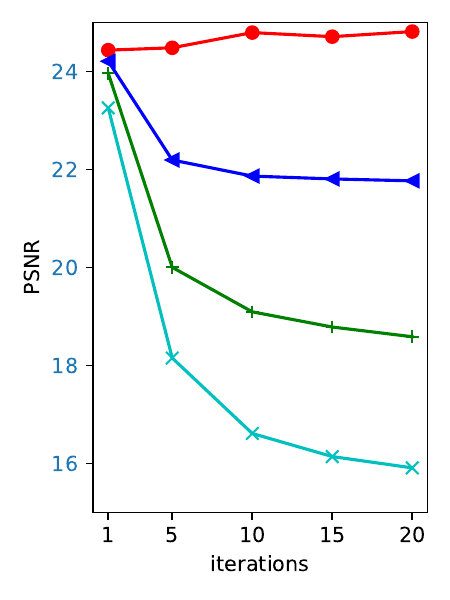}
    
    \end{subfigure}
    \begin{subfigure}[t]{0.19\textwidth}
        \centering
        \includegraphics[width = \textwidth, height=1.2\textwidth]{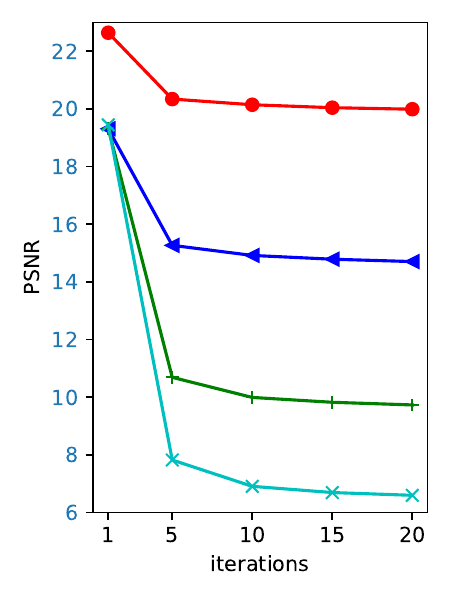}
        
    \end{subfigure}
    \begin{subfigure}[t]{0.19\textwidth}
        \centering
        \includegraphics[width = \textwidth, height=1.2\textwidth]{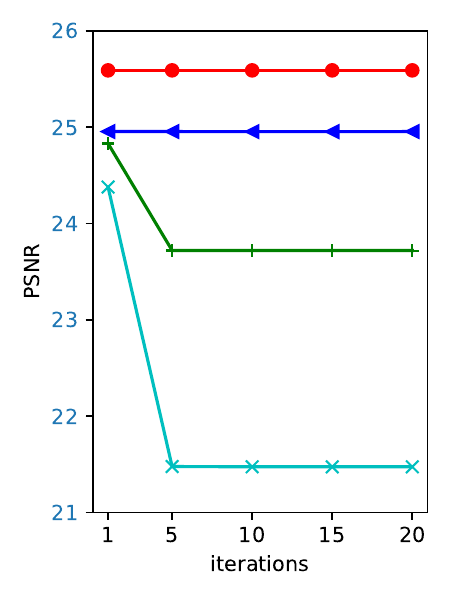}
        
    \end{subfigure}
    \begin{subfigure}[t]{0.19\textwidth}
        \centering
        \includegraphics[width = \textwidth, height=1.2\textwidth]{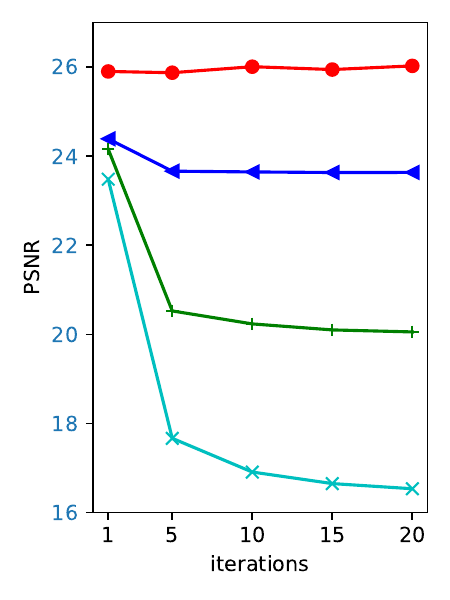}
    
    \end{subfigure}
    \begin{subfigure}[t]{0.19\textwidth}
        \centering
        \includegraphics[width = \textwidth, height=1.2\textwidth]{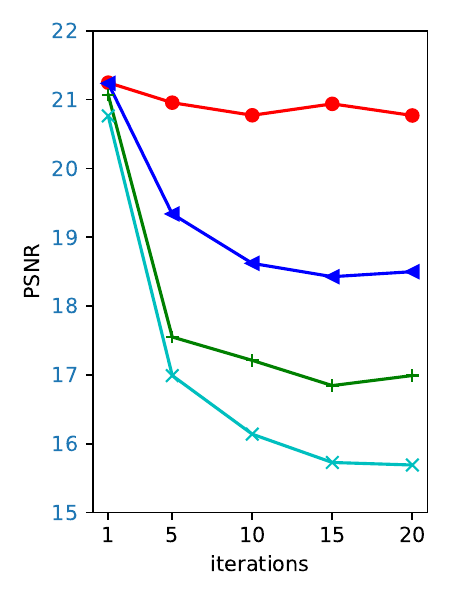}
    
    \end{subfigure}
    
    \begin{subfigure}[t]{0.19\textwidth}
        \centering
        \includegraphics[width = \textwidth, height=1.2\textwidth]{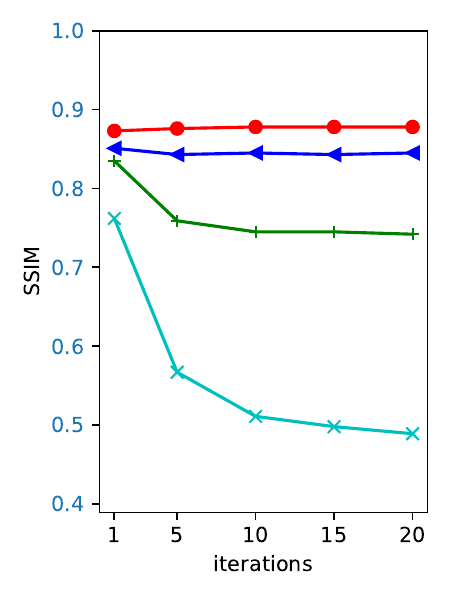}
        \caption{ADMNNet}\label{fig:1f}		
    \end{subfigure}
    \begin{subfigure}[t]{0.19\textwidth}
        \centering
        \includegraphics[width = \textwidth, height=1.2\textwidth]{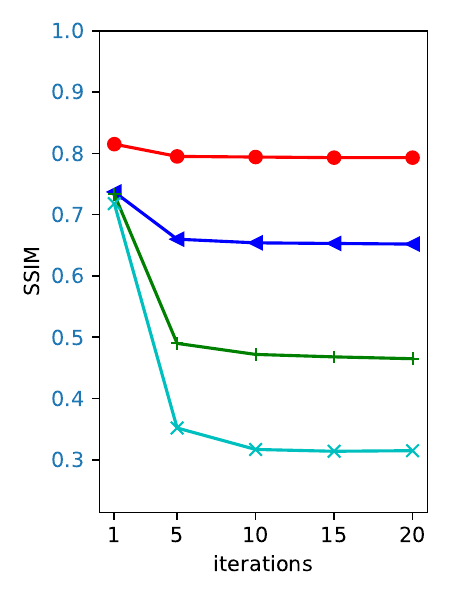}
        \caption{FUnIEGAN}\label{fig:1g}		
    \end{subfigure}
    \begin{subfigure}[t]{0.19\textwidth}
        \centering
        \includegraphics[width = \textwidth, height=1.2\textwidth]{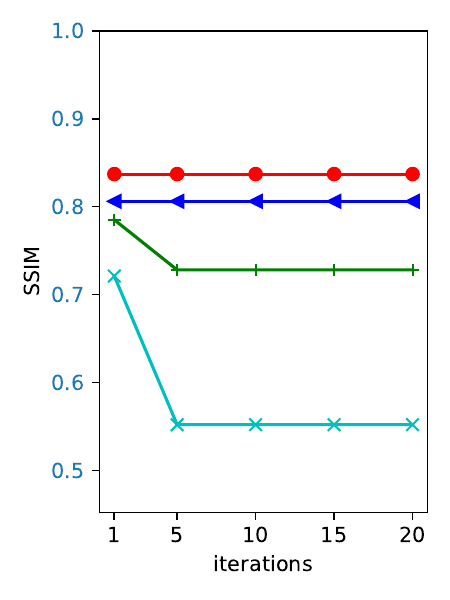}
        \caption{PhysicalNN}\label{fig:1h}		
    \end{subfigure}
    \begin{subfigure}[t]{0.19\textwidth}
        \centering
        \includegraphics[width = \textwidth, height=1.2\textwidth]{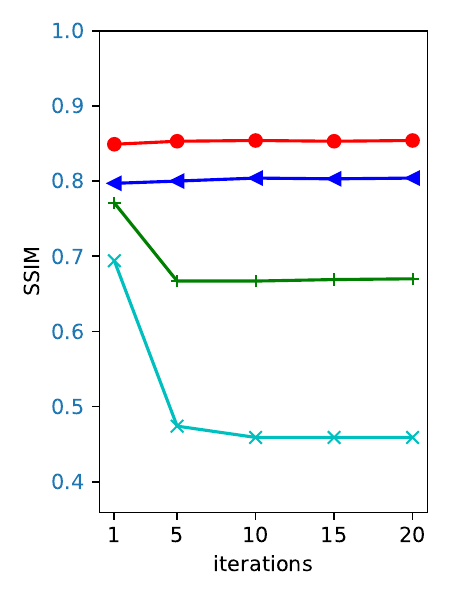}
        \caption{SCNet}\label{fig:1i}		
    \end{subfigure}
    \begin{subfigure}[t]{0.19\textwidth}
        \centering
        \includegraphics[width = \textwidth, height= 1.2\textwidth]{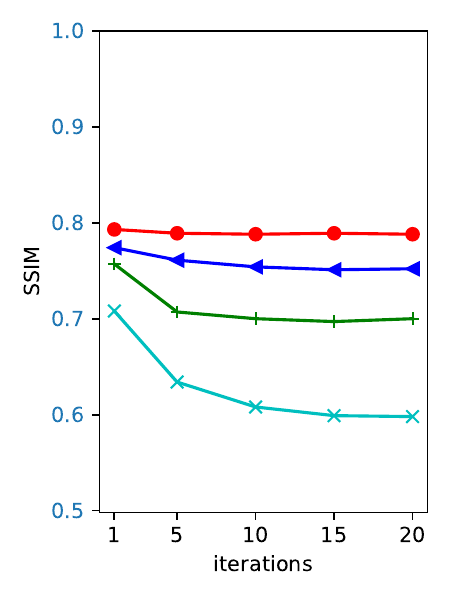}
        \caption{UHD}\label{fig:1j}		
        \end{subfigure}
    \caption{\label{fig:fig1} Quantitive results of adversarial attacks. PSNR and SSIM of undefended UWIE models under adversarial attack on dataset EUVP-I, with $\epsilon\in\{1/255, 2/255, 4/255, 8/255\}$ and iterations $t \in \{1, 5, 10, 15, 20\} $. }
  \end{figure*}
  \begin{figure}[h]
    \centering
    \includegraphics[width = 0.09\textwidth, height =0.09\textwidth]{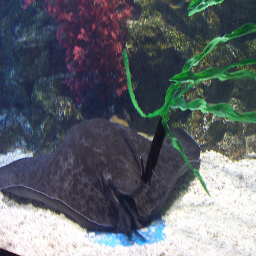}
    \includegraphics[width = 0.09\textwidth, height =0.09\textwidth]{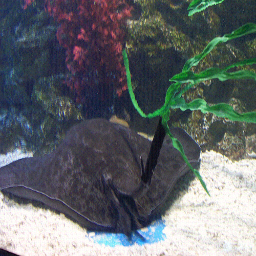}
    \includegraphics[width = 0.09\textwidth, height =0.09\textwidth]{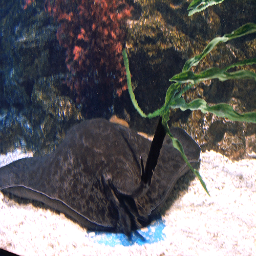}
    \includegraphics[width = 0.09\textwidth, height =0.09\textwidth]{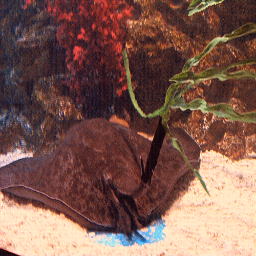}  
    \includegraphics[width = 0.09\textwidth, height =0.09\textwidth]{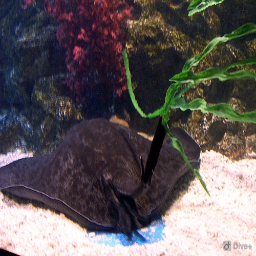}
    \\
    \includegraphics[width = 0.09\textwidth, height =0.09\textwidth]{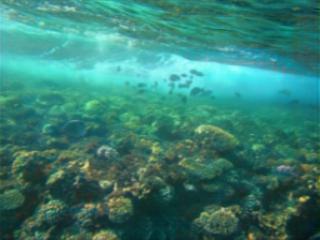}
    \includegraphics[width = 0.09\textwidth, height =0.09\textwidth]{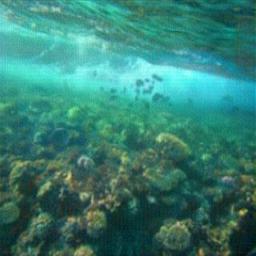}
    \includegraphics[width = 0.09\textwidth, height =0.09\textwidth]{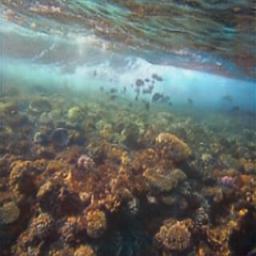}
    \includegraphics[width = 0.09\textwidth, height =0.09\textwidth]{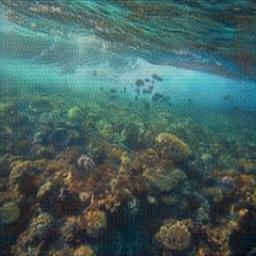}  
    \includegraphics[width = 0.09\textwidth, height =0.09\textwidth]{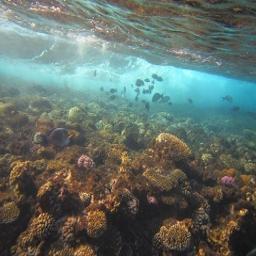}
    \\
    \includegraphics[width = 0.09\textwidth, height =0.09\textwidth]{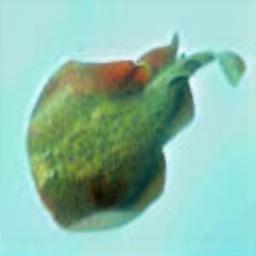}
    \includegraphics[width = 0.09\textwidth, height =0.09\textwidth]{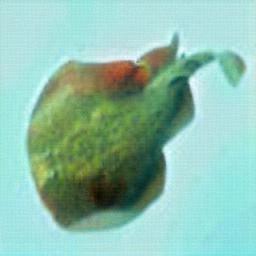}
    \includegraphics[width = 0.09\textwidth, height =0.09\textwidth]{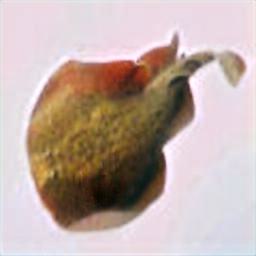}
    \includegraphics[width = 0.09\textwidth, height =0.09\textwidth]{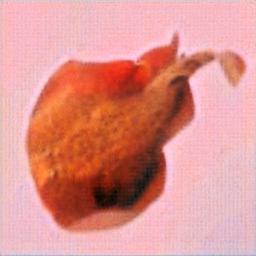}  
    \includegraphics[width = 0.09\textwidth, height =0.09\textwidth]{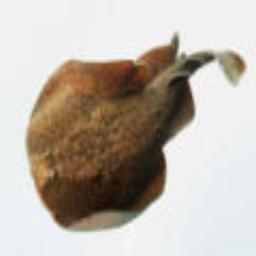}
    \textcolor{black}{\centering{ $x$ \hspace{0.06\textwidth} $x^{\text{adv}}$ \hspace{0.05\textwidth}$f(x)$\hspace{0.05\textwidth} $f(x^{\text{adv}})$\hspace{0.05\textwidth} $y$} }
    \caption{Visual results of adversarial attacks. 
     For each row, the images are degraded input, adversarial examples, 
     original outputs, adversarial outputs of ADMNNet, and ground truth from left to right.}
    \label{fig:2}
    \vspace{-1.7\baselineskip}
  \end{figure}
\subsection{Basic Results of Adversarial Attacks}  
\paragraph{\bf Adversarial Attack Settings.}
If not otherwise specified, adversarial attacks in our paper are based on Pixel Attack.
The selection of relevant parameters for our attack method is diverse. In order to ensure the validity and practicality of experimental results, we set $\epsilon \in \{1/255, 2/255, 4/255, 8/255\}$,  
$ \alpha = 2.0/255$, iterations $ T \in \{1, 5, 10, 15, 20\}$. Perturbations are restricted to $l_{\infty}$ norm.
The parameter selection comprehensively covers reasonable parameter values, making the experimental results more reliable.

We choose the widely used supervised image quality assessment metrics PSNR and SSIM~\cite{wang2004image} to evaluate the effectiveness of our adversarial attacks.
If adversarial examples make the networks generate images that have much lower PSNR and SSIM than those of the original examples, 
they successfully fool the UWIE models. 
In other words, the greater the difference between the metrics of clean example outputs and adversarial example outputs, 
the more vulnerable the model is to adversarial attacks.
\paragraph{ \bf Quantitive results and analysis.}
Table~\ref{tab:1} and Fig.~\ref{fig:fig1} demonstrate the quantitive results of adversarial attacks on five UWIE models. 
All these UWIE models are trained on original examples for 100 epochs. From the results, we can get the following results.

\begin{itemize}
    \item The SSIM and PSNR of all five models exhibit different degrees of decline under the attack of adversarial examples, 
    which indicates that current learning-based UWIE models do have vulnerabilities to adversarial attacks.
    \item A smaller attack intensity ($\epsilon=1/255,2/255$) can hardly have a significant impact 
    on the adversarial robustness of UWIE models.
    The higher the attack intensity ($\epsilon$) and the more iterations of the attack, 
    the more pronounced the effects of adversarial attacks.
    \item Higher attack intensities (increasing the number of iterations from 1 to 5)
    result in the fastest decline in the outputs of various UWIE networks.
    Nevertheless, there exists a lower bound as the iterations of adversarial attacks increase.

    \item 
    The model that performs the best on original examples does not excel on adversarial examples.
    This indicates that adversarial robustness is a new evaluation dimension independent of general image quality assessments.
    \end{itemize}
    \begin{figure*}[t]
        \centering
        \includegraphics[width=9cm,height=0.5cm]{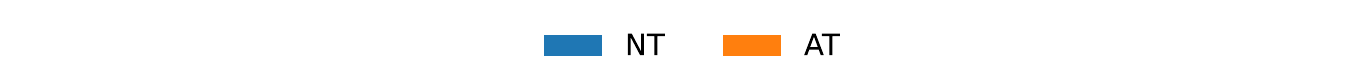}\\
        
        \begin{subfigure}[t]{0.19\textwidth}
            \centering
            \includegraphics[width = \textwidth, height=1.2\textwidth]{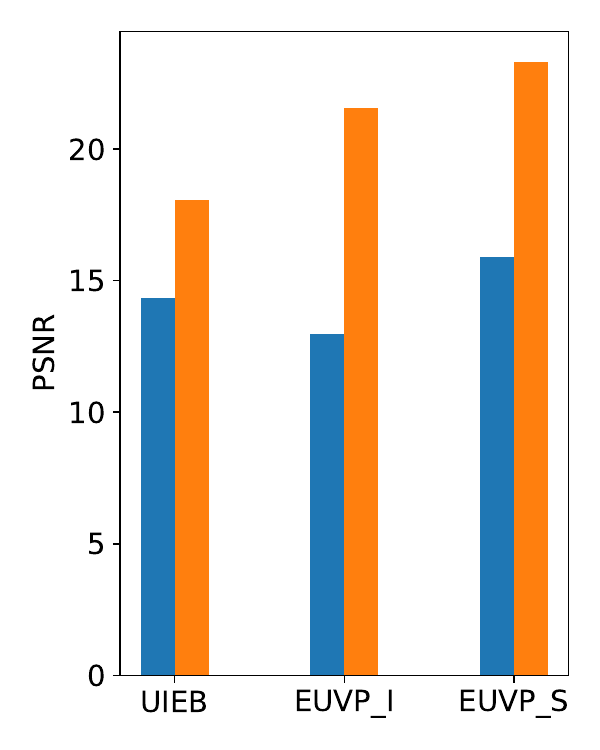}
            
        \end{subfigure}
        \begin{subfigure}[t]{0.19\textwidth}
            \centering
            \includegraphics[width = \textwidth, height=1.2\textwidth]{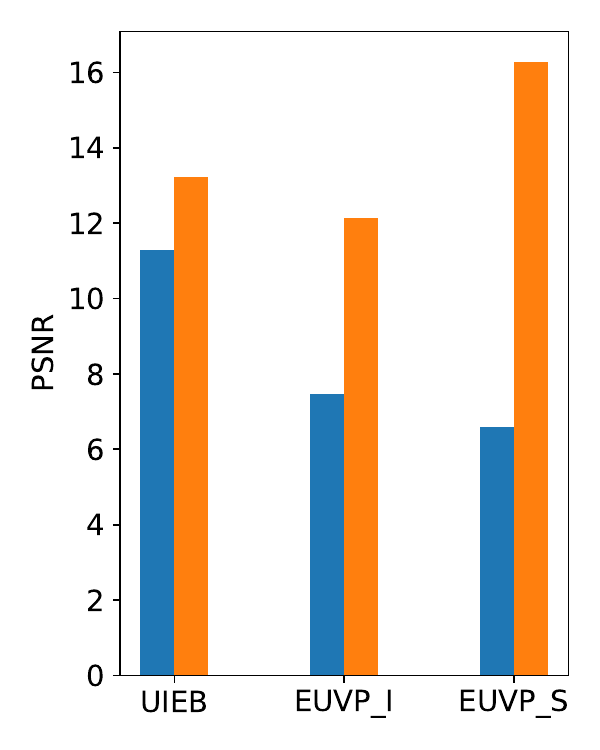}
            
        \end{subfigure}
        \begin{subfigure}[t]{0.19\textwidth}
            \centering
            \includegraphics[width = \textwidth, height=1.2\textwidth]{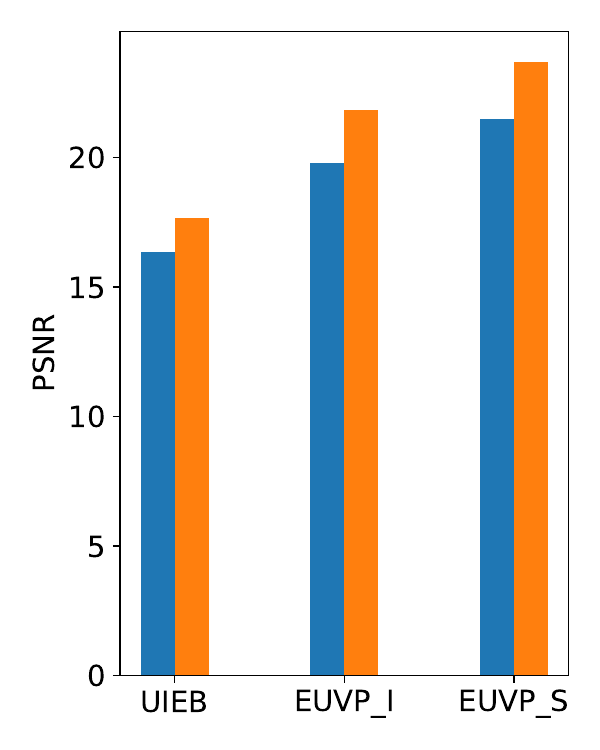}
        
        \end{subfigure}
        \begin{subfigure}[t]{0.19\textwidth}
            \centering
            \includegraphics[width = \textwidth, height=1.2\textwidth]{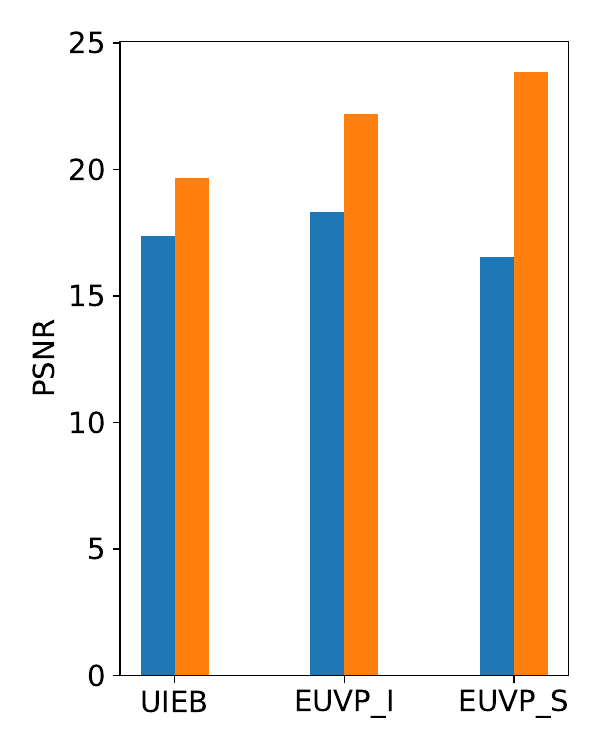}
        
        \end{subfigure}
        \begin{subfigure}[t]{0.19\textwidth}
            \centering
            \includegraphics[width = \textwidth, height=1.2\textwidth]{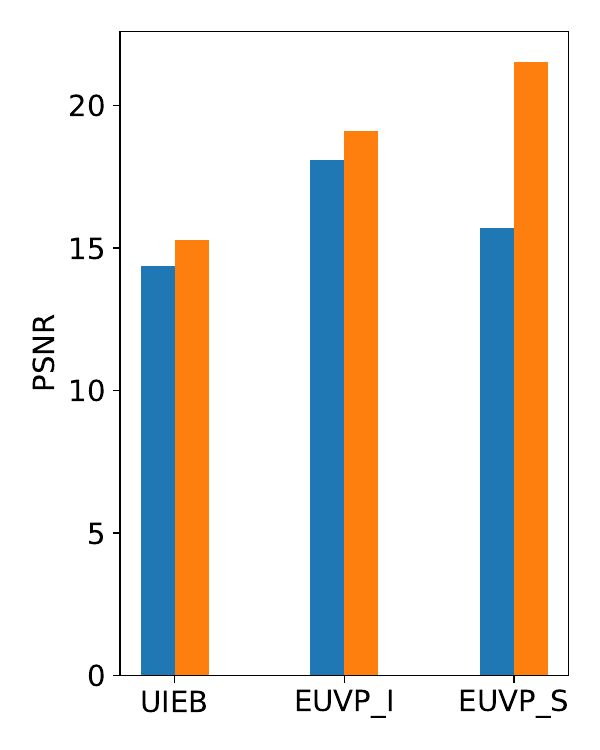}
        
        \end{subfigure}
        
        \begin{subfigure}[t]{0.19\textwidth}
            \centering
            \includegraphics[width = \textwidth, height=1.2\textwidth]{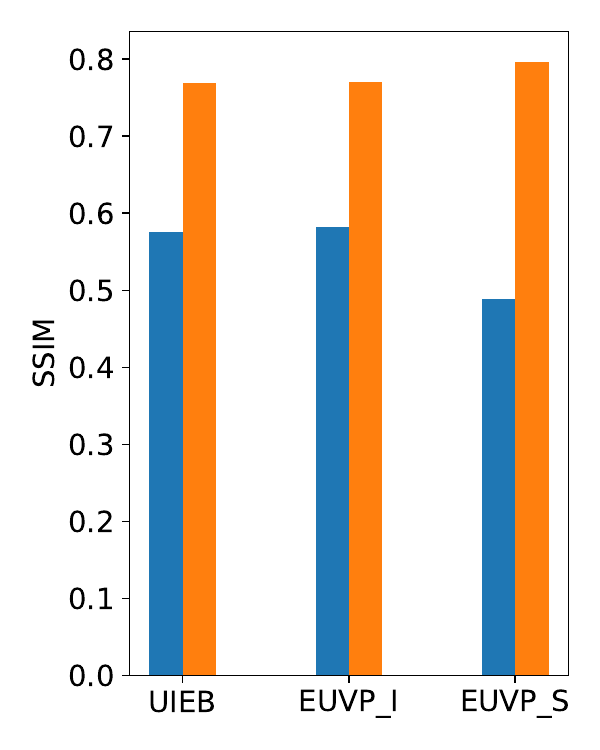}
            \caption{ADMNNet}\label{fig:3f}		
        \end{subfigure}
        \begin{subfigure}[t]{0.19\textwidth}
            \centering
            \includegraphics[width = \textwidth, height=1.2\textwidth]{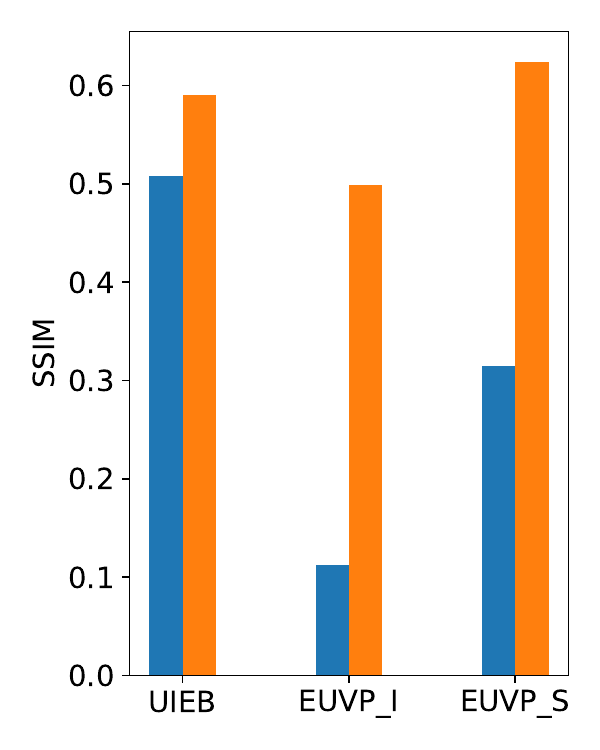}
            \caption{FUnIEGAN}\label{fig:3g}		
        \end{subfigure}
        \begin{subfigure}[t]{0.19\textwidth}
            \centering
            \includegraphics[width = \textwidth, height=1.2\textwidth]{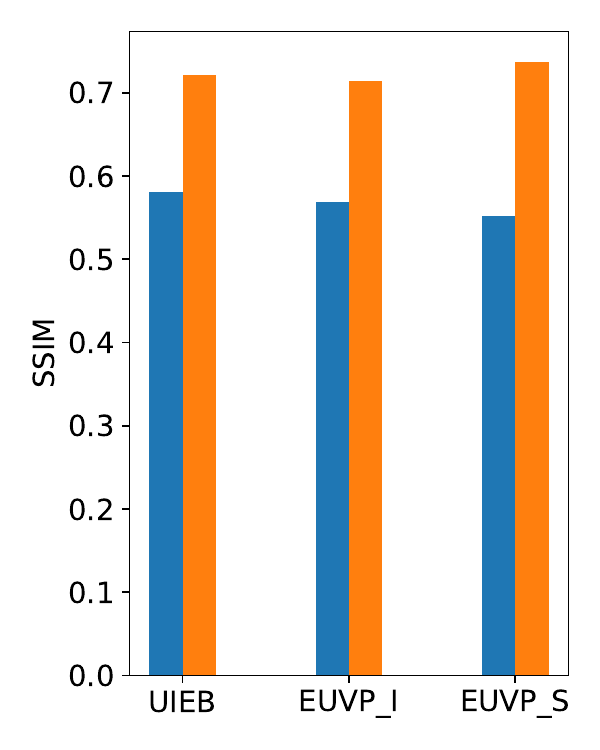}
            \caption{PhysicalNN}\label{fig:3h}		
        \end{subfigure}
        \begin{subfigure}[t]{0.19\textwidth}
            \centering
            \includegraphics[width = \textwidth, height=1.2\textwidth]{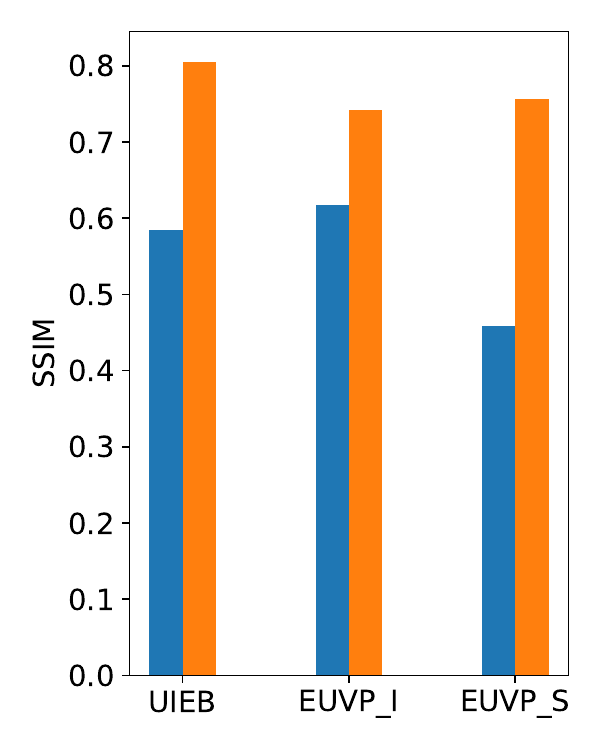}
            \caption{SCNet}\label{fig:3i}		
        \end{subfigure}
        \begin{subfigure}[t]{0.19\textwidth}
            \centering
            \includegraphics[width = \textwidth,height= 1.2\textwidth]{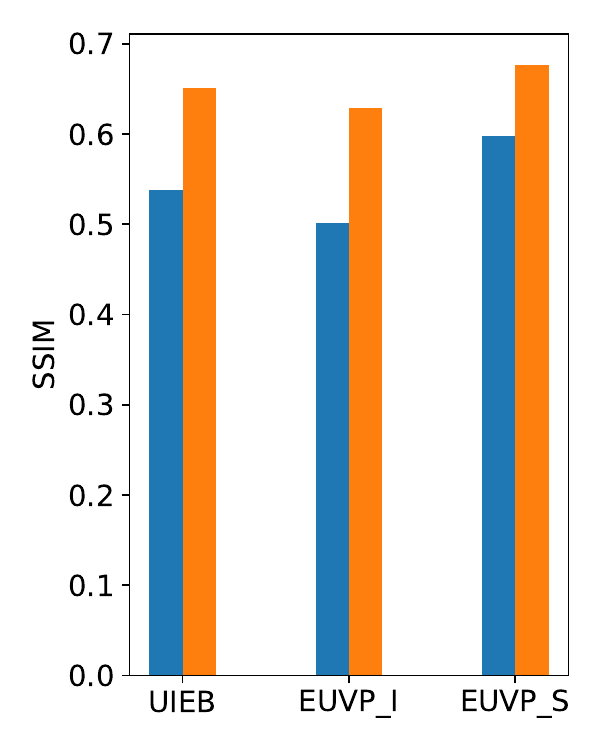}
            \caption{UHD}\label{fig:3j}		
            \end{subfigure}
            \caption{\label{fig:3} Quantitive results of adversarial training. Each column demonstrates the PSNR and SSIM of defended UWIE models. 
            AT represents UWIE models that undergo adversarial training and NT represents those not.}
          \end{figure*}
    \paragraph{\bf Visual results and analysis.}
    Fig.~\ref{fig:2} vividly shows the visual results of adversarial attacks. 
    We can make the following inferences.
 
    \begin{itemize}
        \item Adversarial examples generated by adversarial attack methods can successfully introduce visually imperceptible perturbations,
          aligning with their stealth characteristics.
        \item 
        Adversarial attacks not only lead to a substantial decrease in the output quality of the UWIE model in metrics like PSNR and SSIM, 
        but also have a significant impact on the visual appearance of the output results.
    \end{itemize}

    \paragraph{\bf Summary.}
    Adversarial attacks can dramatically decrease the image quality of learning-based models.
\subsection{Different Adversarial Attack Methods}
To demonstrate the differences between Pixel Attack and Color Shift Attack, We set $\epsilon =8/255$,  
$ \alpha = 2.0/255$, iterations $ t =20$. Perturbations are restricted to $l_{\infty}$ norm.

\begin{figure}[t]
    \centering
    \includegraphics[width = 0.09\textwidth, height =0.09\textwidth]{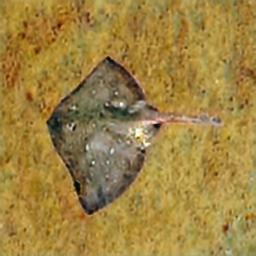}
    \includegraphics[width = 0.09\textwidth, height =0.09\textwidth]{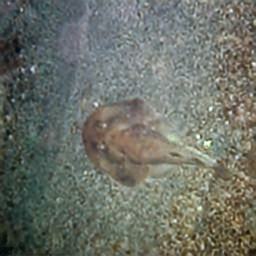}  
    \includegraphics[width = 0.09\textwidth, height =0.09\textwidth]{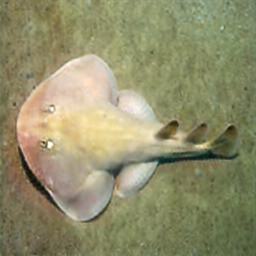}
    \includegraphics[width = 0.09\textwidth, height =0.09\textwidth]{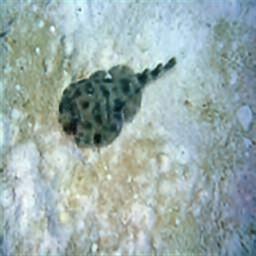}  
    \includegraphics[width = 0.09\textwidth, height =0.09\textwidth]{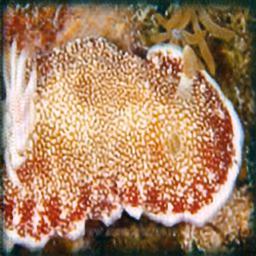}\\
    \includegraphics[width = 0.09\textwidth, height =0.09\textwidth]{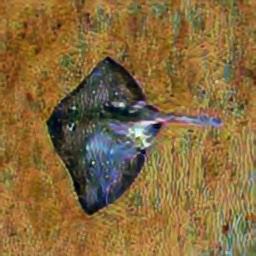}
    \includegraphics[width = 0.09\textwidth, height =0.09\textwidth]{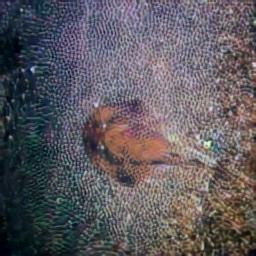}  
    \includegraphics[width = 0.09\textwidth, height =0.09\textwidth]{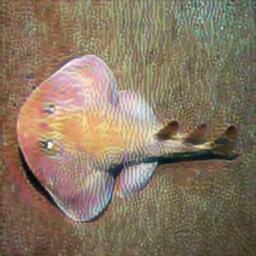}
    \includegraphics[width = 0.09\textwidth, height =0.09\textwidth]{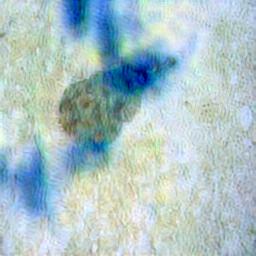}  
    \includegraphics[width = 0.09\textwidth, height =0.09\textwidth]{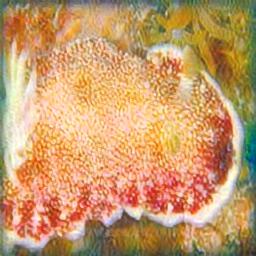}\\
    \includegraphics[width = 0.09\textwidth, height =0.09\textwidth]{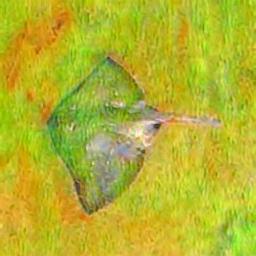}
    \includegraphics[width = 0.09\textwidth, height =0.09\textwidth]{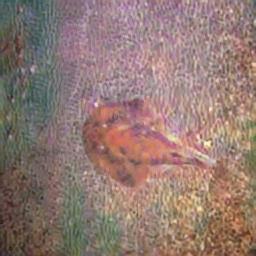}  
    \includegraphics[width = 0.09\textwidth, height =0.09\textwidth]{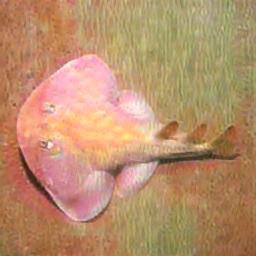}
    \includegraphics[width = 0.09\textwidth, height =0.09\textwidth]{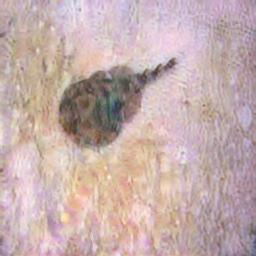}  
    \includegraphics[width = 0.09\textwidth, height =0.09\textwidth]{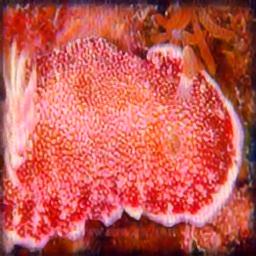}
    \caption{Visual results of two different adversarial attacks. Each row represents the original outputs, the adversarial outputs of Pixel Attack, and the adversarial outputs of Color Shift Attack from up to down. Each column represents different instances.}
    \vspace{-\baselineskip}
    \label{fig:6}
  \end{figure}

\paragraph{\bf Visual Results and analysis.}
The visual results of adversarial examples generated by Pixel Attack and Color Shift Attack are shown in Fig.~\ref{fig:6}.
Compared with original outputs, Pixel Attack makes adversarial outputs a failure of image restoration,
 resulting in illumination changes, and sometimes artifacts in the image.
Color Shift Attack, on the other hand, makes the adversarial outputs have a color shift and has little impact on the image contents.
\paragraph{\bf Quantitive Results and analysis.}
To quantitatively evaluate the effectiveness of different adversarial attacks, we calculate the PNSR between adversarial outputs and ground truth.
Apart from Pixel Attack and Color Shift Attack, we perform extra adversarial attacks on RGB channels respectively. The channel-specific adversarial attacks are expressed as follows,
\begin{equation}
    x^{0} = x+ U(-\epsilon,\epsilon) \odot M,
    \label{eq:9}
\end{equation}
\begin{equation}
    x^{t+1} = \Pi_{\mathcal{B}_\infty(x,\epsilon)}(x^{t}+\alpha M\odot \text{sgn}(\nabla_{x^{t}} \mathcal{L}(f(x^{t}),y))) ,
    \label{eq:10}
\end{equation}
where $\odot$ is the Hadamard product. $M$ is a mask that only contains one of the RGB channels.
Table~\ref{tab:my-table} shows the PSNR of adversarial outputs under different adversarial attacks, 
which indicates that Pixel Attack and Color Shift Attack have the best performance at the same level of adversarial attack strength.
However, simply attacking a specific RGB channel does not effectively produce a good attack effect on the UWIE model.
\subsection{Adversarial Defenses to UWIE Models}
  \begin{table}[]
    \centering
    \setlength{\tabcolsep}{0.27cm}
    \caption{PSNR ($\uparrow$) of Adversarial outputs under Pixel Attack, Color Attack, and Single Channel perturbations.}
    \label{tab:my-table}
    \begin{tabular*}{\linewidth}{@{}rrrrrr@{}}
    \toprule
               & Pixel    & Color  & R      & G      & B      \\ \midrule
    ADMNNet    & 13.000 & \textbf{12.703} & 18.780 & 15.243 & 19.774 \\
    FUnIEGAN   & \textbf{7.496}  & 10.928 & 13.707 & 11.515 & 13.726 \\
    PhysicalNN & \textbf{19.943} & 20.954 & 21.827 & 22.286 & 22.450 \\
    UHD        & 18.102 & \textbf{18.029} & 19.321 & 19.786 & 20.014 \\
    SCNet      & \textbf{16.618} & 18.623 & 19.718 & 19.745 & 22.087 \\ \bottomrule
    \end{tabular*}
    \end{table}

    \begin{figure*}[t]
        \centering
        \begin{subfigure}[t]{0.19\textwidth}
            \centering
            \includegraphics[width = \textwidth, height=1.2\textwidth]{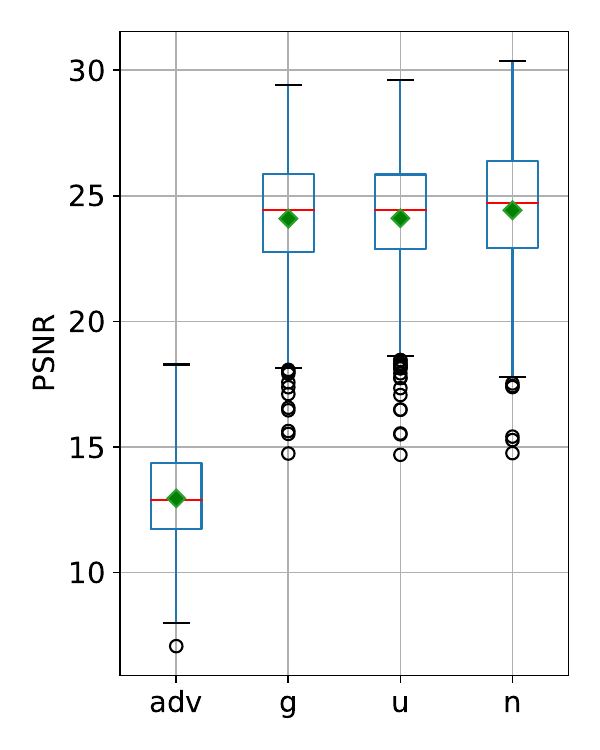}
        \end{subfigure}
        \begin{subfigure}[t]{0.19\textwidth}
            \centering
            \includegraphics[width = \textwidth, height=1.2\textwidth]{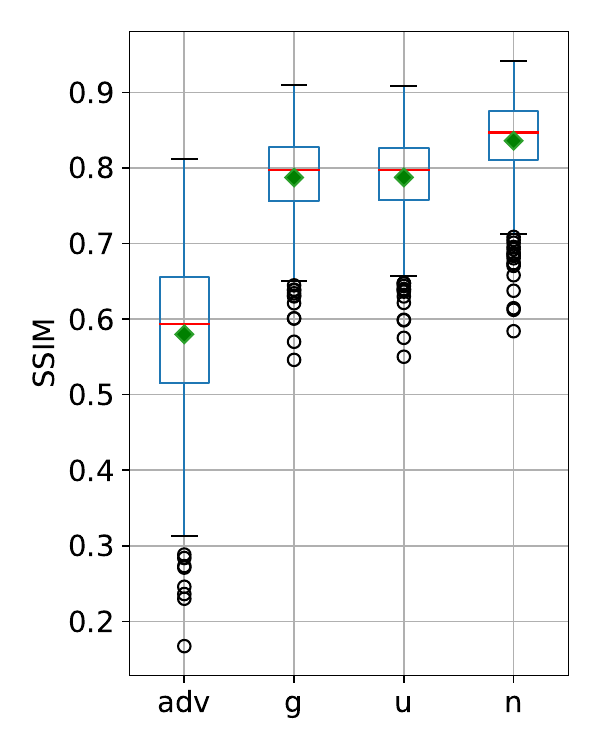}
        \end{subfigure}
        \begin{subfigure}[t]{0.19\textwidth}
            \centering
            \includegraphics[width = \textwidth, height=1.2\textwidth]{figures/noise/EUVP_I/ADMNNetPSNR.pdf}
        \end{subfigure}
        \begin{subfigure}[t]{0.19\textwidth}
            \centering
            \includegraphics[width = \textwidth, height=1.2\textwidth]{figures/noise/EUVP_I/ADMNNetSSIM.pdf}
        \end{subfigure}
        \begin{subfigure}[t]{0.19\textwidth}
            \centering
            \includegraphics[width = \textwidth, height=1.2\textwidth]{figures/noise/EUVP_I/ADMNNetPSNR.pdf}
        \end{subfigure}
        \begin{subfigure}[t]{0.19\textwidth}
            \centering
            \includegraphics[width = \textwidth, height=1.2\textwidth]{figures/noise/EUVP_I/ADMNNetSSIM.pdf}
            \caption{ADMNNet}\label{fig:5f}		
        \end{subfigure}
        \begin{subfigure}[t]{0.19\textwidth}
            \centering
            \includegraphics[width = \textwidth, height=1.2\textwidth]{figures/noise/EUVP_I/ADMNNetPSNR.pdf}
            \caption{FUnIEGAN}\label{fig:5g}
        \end{subfigure}
        \begin{subfigure}[t]{0.19\textwidth}
            \centering
            \includegraphics[width = \textwidth, height=1.2\textwidth]{figures/noise/EUVP_I/ADMNNetSSIM.pdf}
            \caption{PhysicalNN}\label{fig:5h}		
        \end{subfigure}
        \begin{subfigure}[t]{0.19\textwidth}
            \centering
            \includegraphics[width = \textwidth, height=1.2\textwidth]{figures/noise/EUVP_I/ADMNNetPSNR.pdf}
            \caption{SCNet}\label{fig:5i}		
        \end{subfigure}
        \begin{subfigure}[t]{0.19\textwidth}
            \centering
            \includegraphics[width = \textwidth, height= 1.2\textwidth]{figures/noise/EUVP_I/ADMNNetSSIM.pdf}
            \caption{UHD}\label{fig:5j}		
            \end{subfigure}
            \caption{\label{fig:fig5} Boxplots of adversarial perturbations and random noise. Adversarial perturbations, Gaussian noise, uniform noise and no noise are denoted as adv, g, u, n.}
          \end{figure*}

\paragraph{\bf Adversarial training settings.}
We perform adversarial training both on pre-trained models and from scratch. For pre-trained models, we conduct adversarial training for 20 epochs.
All hyperparameters remain unchanged from the pre-training, apart from the training loss. 
The training loss is modified to the sum of the clean loss and adversarial training regularization term.
For models trained from scratch, we perform adversarial training for 100 epochs. All the hyperparameters remain the same with pre-training settings.
As mentioned above, we define the adversarial training regularization term as the $l_{2}$ norm between adversarial examples and original examples, 
which aims to narrow the gap between original examples and adversarial examples~\cite{zhang2019theoretically}.
In the adversarial training phase, the regularization coefficient $\gamma$ is fixed to 1 for all models. 
Adversarial attack parameters are set to $\epsilon \in \{4/255,8/255\}$ and $ \alpha = 2.0$, iterations $ t =20$. Adversarial loss for the attack method is MSE loss.
\begin{figure}[t]
    \centering
    \includegraphics[width = 0.09\textwidth, height =0.09\textwidth]{figures/demo/918_img_.png/raw.png}
    \includegraphics[width = 0.09\textwidth, height =0.09\textwidth]{figures/demo/918_img_.png/not_defend_ADMNNet.png}  
    \includegraphics[width = 0.09\textwidth, height =0.09\textwidth]{figures/demo/918_img_.png/not_adv_out.png}
    \includegraphics[width = 0.09\textwidth, height =0.09\textwidth]{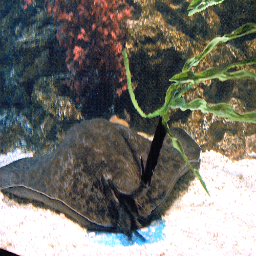}  
    \includegraphics[width = 0.09\textwidth, height =0.09\textwidth]{figures/demo/918_img_.png/gt.png}
  \\
    \includegraphics[width = 0.09\textwidth, height =0.09\textwidth]{figures/demo/im_f12_.jpg/raw.png}
    \includegraphics[width = 0.09\textwidth, height =0.09\textwidth]{figures/demo/im_f12_.jpg/not_defend_ADMNNet.png}  
    \includegraphics[width = 0.09\textwidth, height =0.09\textwidth]{figures/demo/im_f12_.jpg/not_adv_out.png}
    \includegraphics[width = 0.09\textwidth, height =0.09\textwidth]{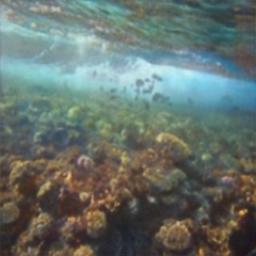}  
    \includegraphics[width = 0.09\textwidth, height =0.09\textwidth]{figures/demo/im_f12_.jpg/gt.png}
    \\
    \includegraphics[width = 0.09\textwidth, height =0.09\textwidth]{figures/demo/n01496331_3304.jpg/raw.png}
    \includegraphics[width = 0.09\textwidth, height =0.09\textwidth]{figures/demo/n01496331_3304.jpg/not_defend_ADMNNet.png}  
    \includegraphics[width = 0.09\textwidth, height =0.09\textwidth]{figures/demo/n01496331_3304.jpg/not_adv_out.png}
    \includegraphics[width = 0.09\textwidth, height =0.09\textwidth]{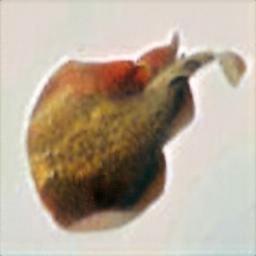}  
    \includegraphics[width = 0.09\textwidth, height =0.09\textwidth]{figures/demo/n01496331_3304.jpg/gt.png}
    \textcolor{black}{\centering{ $x$ \hspace{0.05\textwidth} $f(x^{\text{adv}})$ \hspace{0.04\textwidth}$f_{\text{d}}(x)$\hspace{0.03\textwidth} $f_{\text{d}}(x^{\text{adv}})$\hspace{0.04\textwidth} $y$} }
    \caption{ Visual results of adversarial training. For each row, the images are degraded input, adversarial results, 
     original outputs, defended output for ADMNNet, and ground truth. $f_\text{d}(\cdot)$ represents the adversarial trained UWIE models.}
    \label{fig:7}

  \end{figure}
  \begin{figure*}[t]
    \centering
    \begin{subfigure}[t]{0.15\textwidth}
        \centering
        \includegraphics[height = \textwidth, width =1.0\textwidth ]{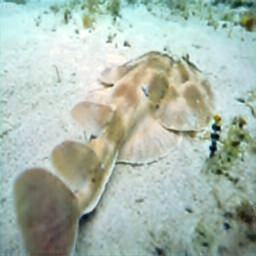}		
    \end{subfigure}
    \begin{subfigure}[t]{0.15\textwidth}
        \centering
        \includegraphics[height = \textwidth, width =1.0\textwidth ]{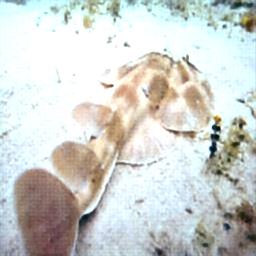}		
    \end{subfigure}
    \begin{subfigure}[t]{0.15\textwidth}
        \centering
        \includegraphics[height = \textwidth, width =1.0\textwidth ]{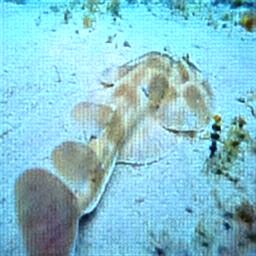}		
    \end{subfigure}
    \begin{subfigure}[t]{0.15\textwidth}
        \centering
        \includegraphics[height = \textwidth, width =1.0\textwidth ]{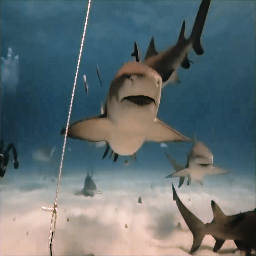}		
    \end{subfigure}
    \begin{subfigure}[t]{0.15\textwidth}
        \centering
        \includegraphics[height = \textwidth, width =1.0\textwidth ]{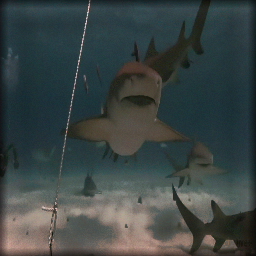}		
    \end{subfigure}
    \begin{subfigure}[t]{0.15\textwidth}
        \centering
        \includegraphics[height = \textwidth, width =1.0\textwidth ]{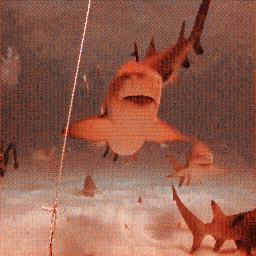}
    \end{subfigure}\\
    \centering
    \begin{subfigure}[t]{0.15\textwidth}
        \centering
        \includegraphics[height = \textwidth, width =1.0\textwidth ]{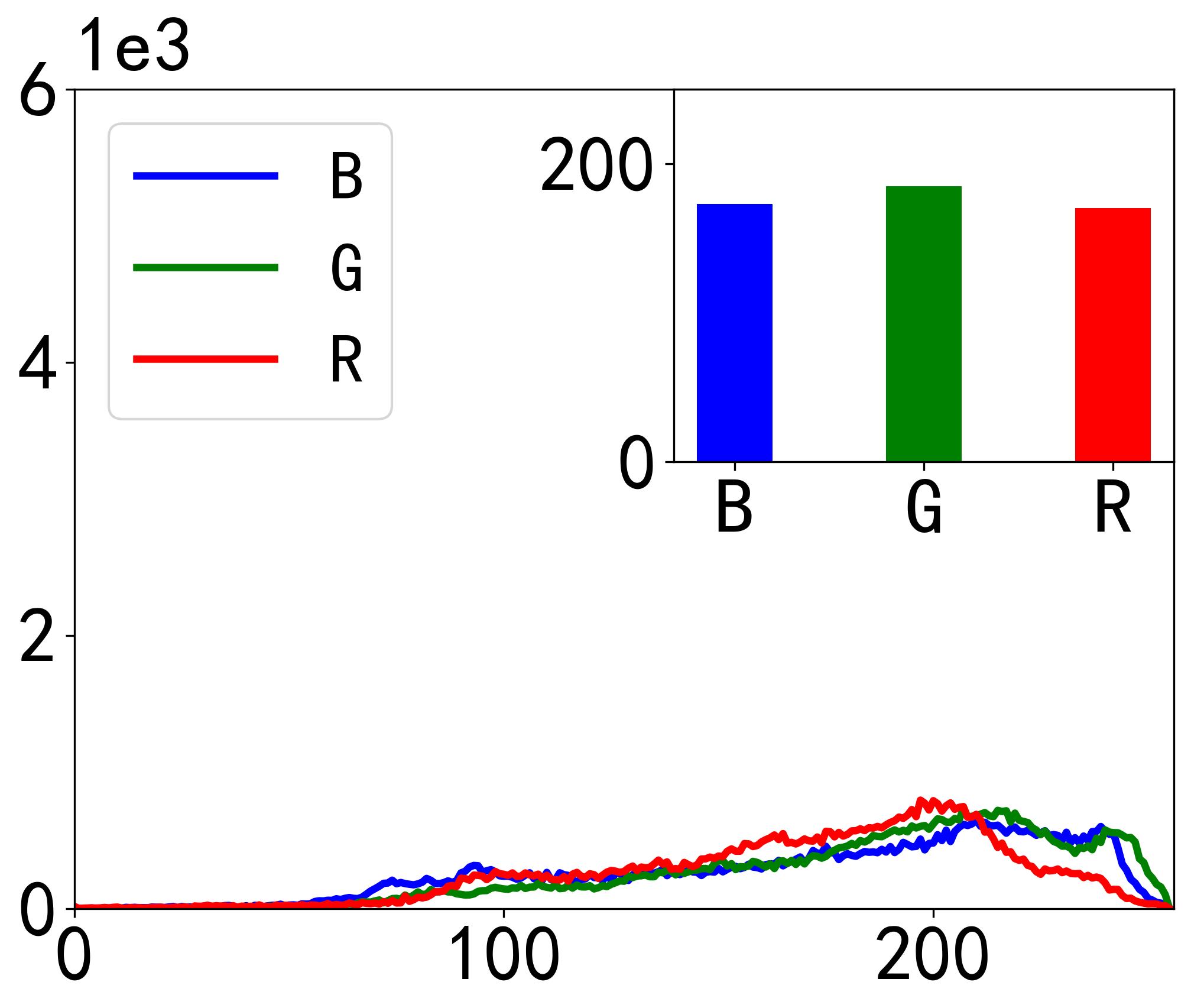}		
    \end{subfigure}
    \begin{subfigure}[t]{0.15\textwidth}
        \centering
        \includegraphics[height = \textwidth, width =1.0\textwidth ]{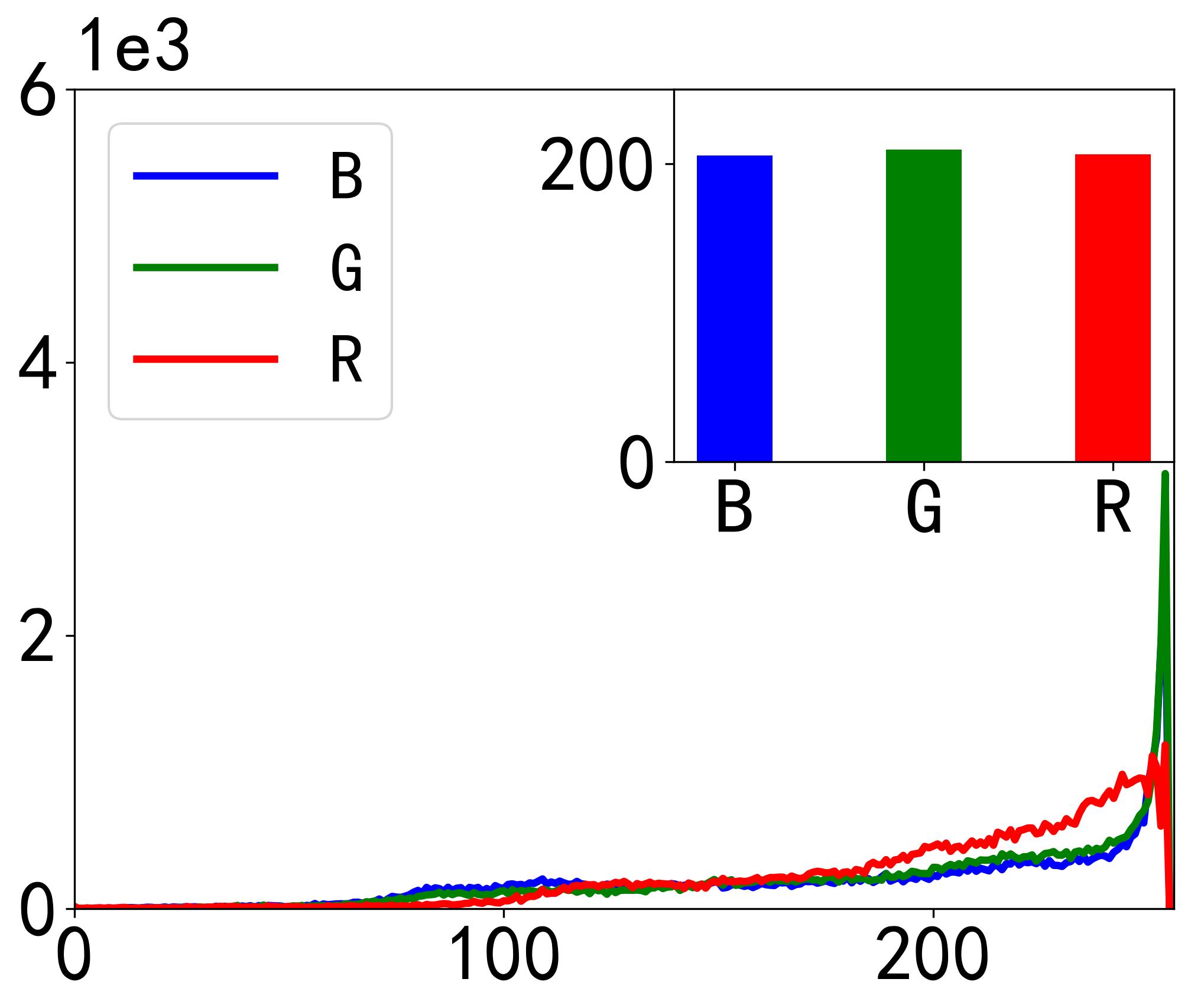}		
    \end{subfigure}
    \begin{subfigure}[t]{0.15\textwidth}
        \centering
        \includegraphics[height = \textwidth, width =1.0\textwidth ]{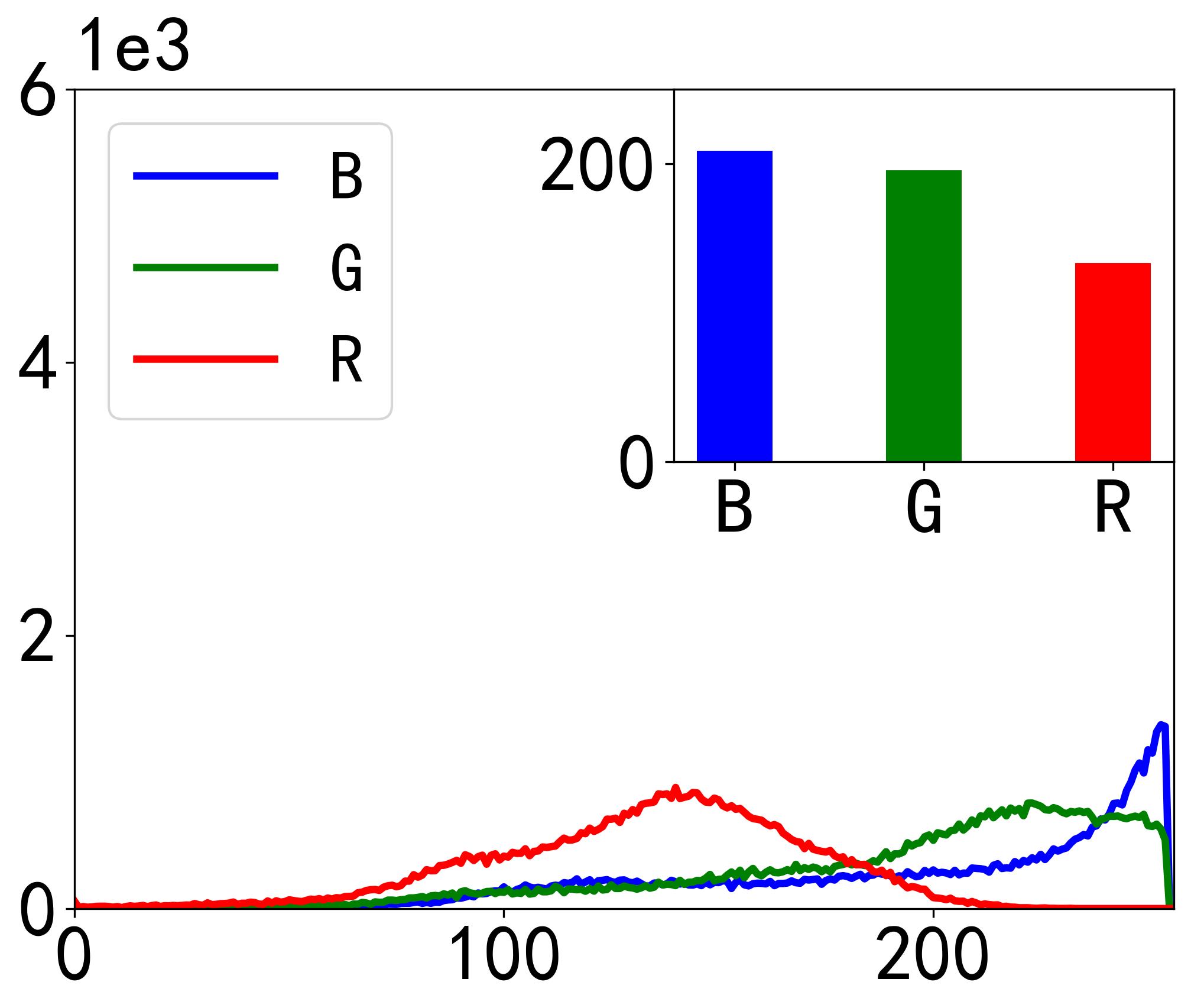}		
    \end{subfigure}
    \begin{subfigure}[t]{0.15\textwidth}
        \centering
        \includegraphics[height = \textwidth, width =1.0\textwidth ]{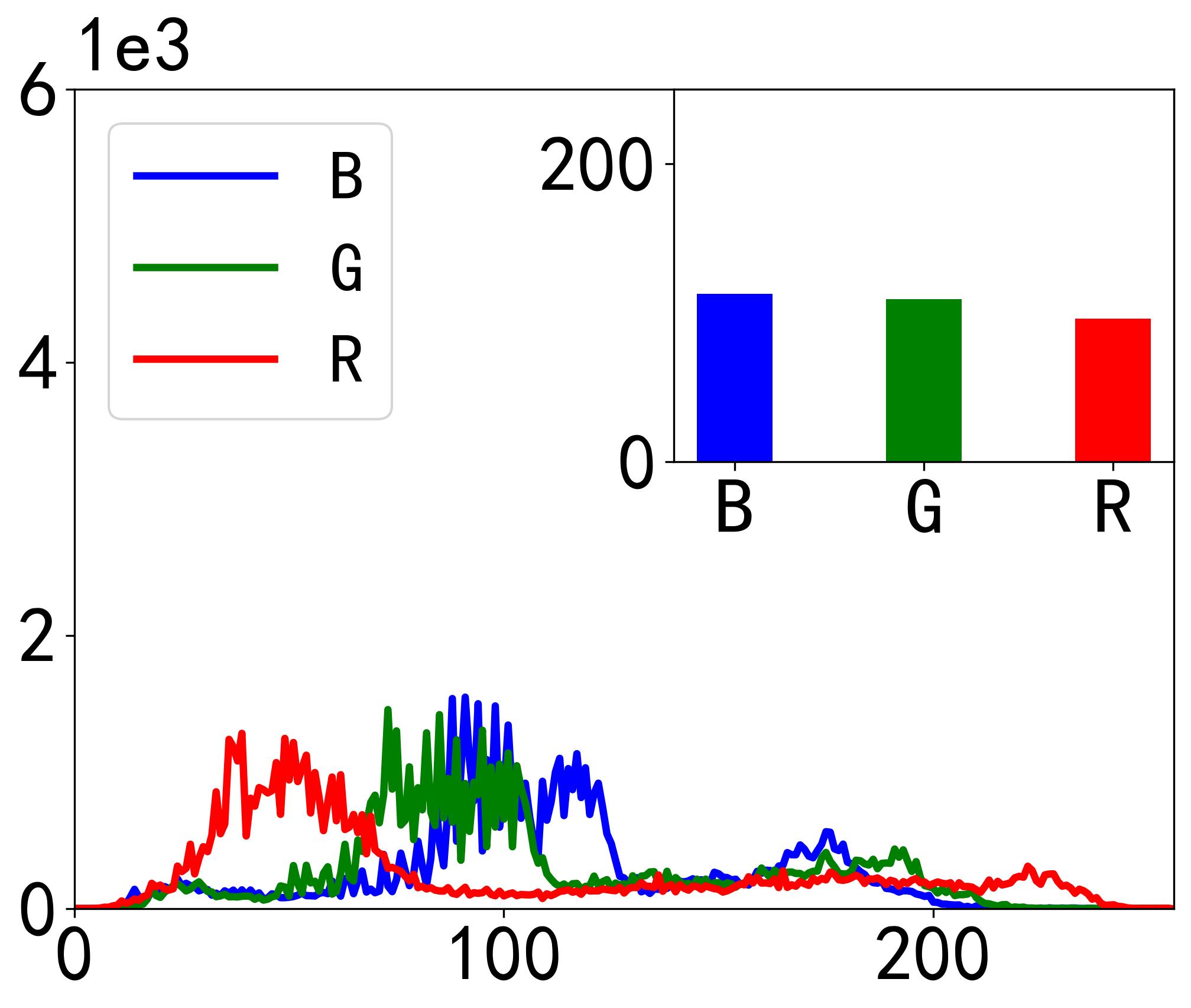}		
    \end{subfigure}
    \begin{subfigure}[t]{0.15\textwidth}
        \centering
        \includegraphics[height = \textwidth, width =1.0\textwidth ]{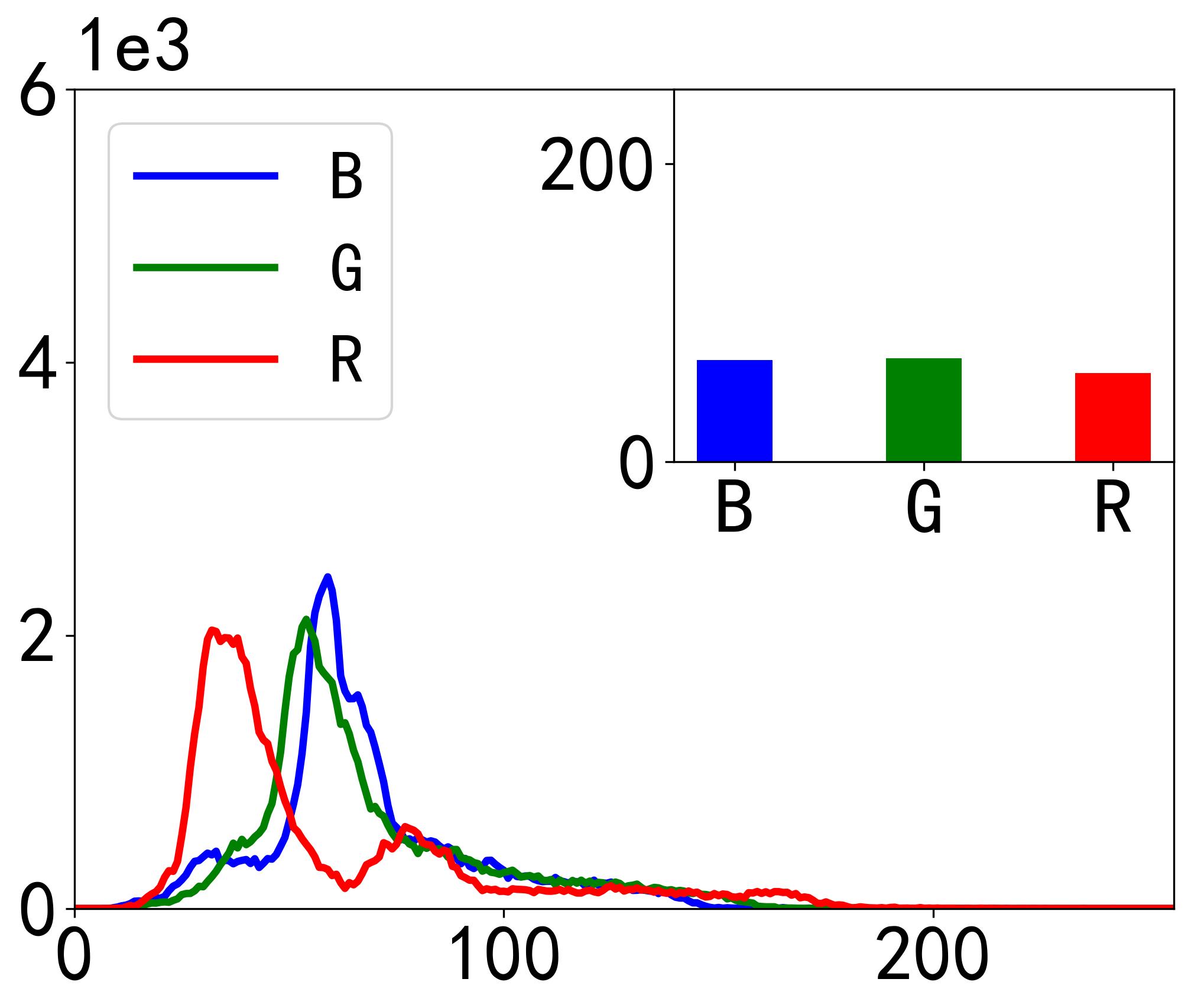}		
    \end{subfigure}
    \begin{subfigure}[t]{0.15\textwidth}
        \centering
        \includegraphics[height = \textwidth, width =1.0\textwidth ]{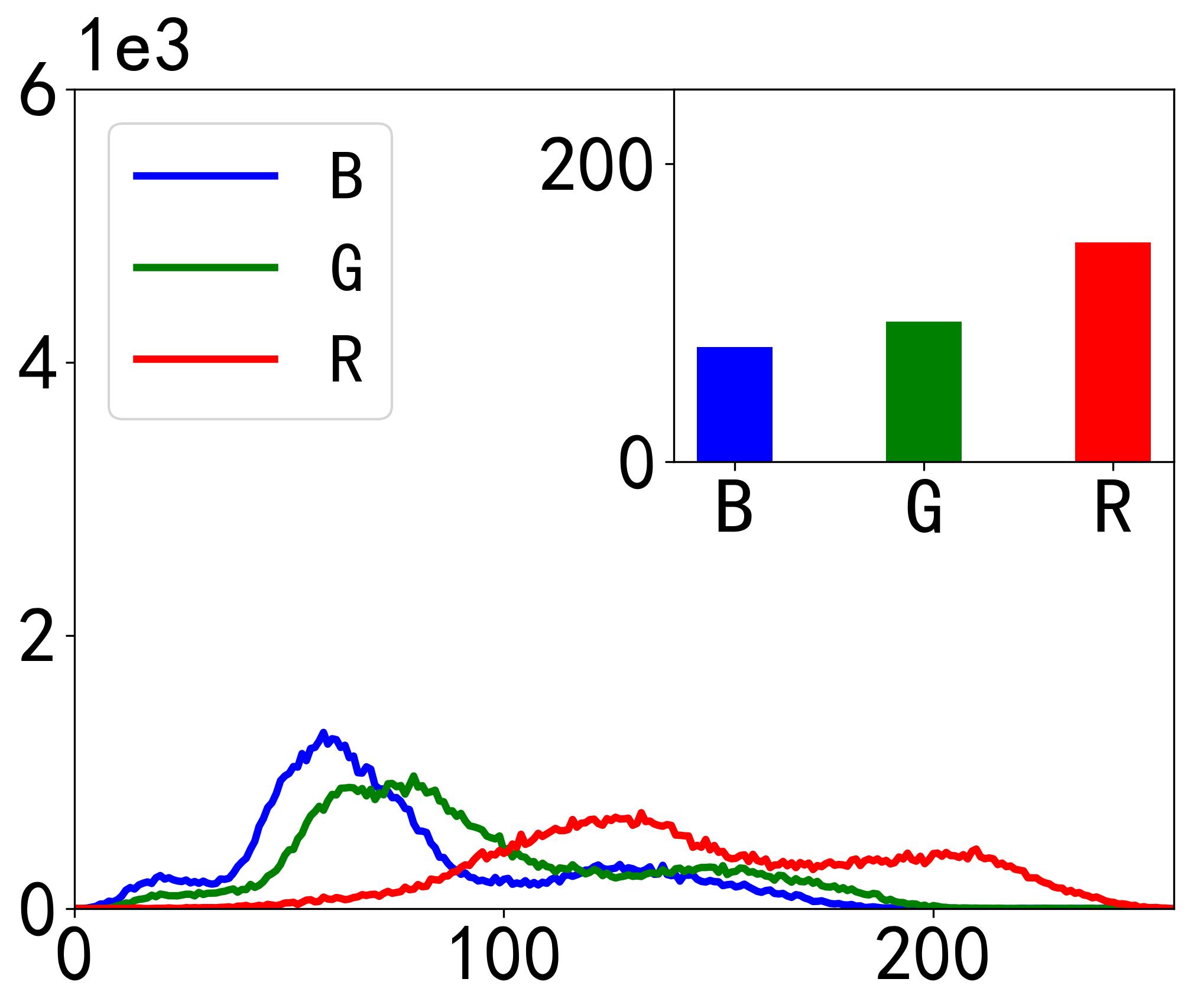}
    \end{subfigure}
    \caption{Histogram analysis of adversarial outputs. Two groups of images are demonstrated. For each group, the images are original outputs, pixel attack outputs, and color attack outputs from left to right.}
\end{figure*}
\paragraph{\bf Quantitive results.}
The adversarial attack conducted on defended and undefended UWIE models is parameterized with  $\alpha = 2/255, \epsilon=8/255,T = 20$.
As shown in Fig.~\ref{fig:3}, we can get the following results.
\begin{itemize}
    \item Even the models that are merely finetuned by adversarial training for 20 epochs show considerate adversarial robustness compared with models without adversarial training, which indicates that adversarial training is a effective defense tragedy.
    \item For the most effective adversarial training, the PSNR remarkably increases by 9.68 from 6.59 to 16.27 and the SSIM increases by 0.315 from 0.31 to 0.625 (FunIEGAN on EUVP-S).
\end{itemize}

\paragraph{\bf Visual results.}
    Fig.~\ref{fig:7} demonstrates the results of adversarial training. We get the following results,
    \begin{itemize}
        \item The enhanced results of adversarial examples and original examples become more consistent after adversarial training,
        effectively mitigating issues such as color distortion and changes in illumination caused by adversarial examples, 
        which indicates that our adversarial training method is effective in defending learning-based UWIE models from adversarial attacks.
        \item The adversarial training makes the UWIE models resist the threat of adversarial exampless and does not compromise the performance of original examples. 
        $f(x)$ in Fig.~\ref{fig:2} and $f_{\text{d}}(x)$ in Fig.~\ref{fig:7} demonstrate that adversarial trained models function well both on the adversarial and original examples.
    \end{itemize}
   
\paragraph{\bf Summary.}
In summary, adversarial training is capable of defending learning-based UWIE models. The models with adversarial training generate more appealing enhanced results than those without any defending strategies.

\subsection{Further Discussions}
\paragraph{\bf Random noise and adversarial perturbations.}
To further study the effectiveness of adversarial noise, 
we evaluate the impact of Gaussian noise, uniform noise, and adversarial perturbations on the outputs of the undefended UWIE model under the same noise intensity.
The Gaussian noise has a mean of 0 and a variance of $\epsilon$, the uniform noise is distributed in the range $[-\epsilon, \epsilon]$, and the adversarial attack is Pixel attack with $\epsilon = 8/255$.

Fig.~\ref{fig:fig5} shows the boxplots of SSIM and PSNR of all five UWIE models on dataset EUVP-I. 
It is evident from Fig.~\ref{fig:fig5} (a)-(j) that the random noise of the same intensity, whether Gaussian or uniform, has only a slight impact on the output metrics of the UWIE models.
In contrast, adversarial perturbations have a significant impact on SSIM and PSNR values of UWIE models, which indicate that adversarial perturbations have apparent differences from common random perturbations.
Research on defending against adversarial attacks on UWIE models is of great significance.

\begin{table}[t]
    \centering
    \setlength{\tabcolsep}{0.44cm}
    \caption{PSNR between $x$ and $x^{\text{adv}}$ on of all discussed UWIE models on dataset
    EUVP-I. The first column represents the adversarial attack parameter $\epsilon$ (/255). (A)- (E) represents ADMNNet, FUnIEGAN, PhysicalNN, UHD, and SCNet respectively.}
    \label{tab:imp}
    \begin{tabular*}{\linewidth}{@{}llllll@{}}
    \toprule
     $\epsilon$ & (A) & (B) & (C) & (D)   & (E) \\ \midrule
    1 & 46.90   & 46.88    & 46.80      & 46.86 & 46.89 \\
    2 & 41.46   & 40.87    & 40.71      & 40.79 & 41.01 \\
    4 & 36.20   & 34.90    & 34.55      & 34.81 & 35.31 \\
    8 & 31.18   & 28.98    & 28.31      & 28.70 & 29.75 \\ \bottomrule
    \end{tabular*}
    \end{table}
\paragraph{\bf Histogram analysis}
To further study the differences between Pixel Attack and Color Shift Attack, we conduct a histogram analysis of the adversarial outputs of the two attacks. 
In Fig.~\ref{fig:7}, Pixel Attack significantly changes the mean value of the outputs, and seems to make each channel get closer, which makes the outputs overexposure or underexposure,.
Yet Color Shift Attack makes a diversified distribution of pixel values for each channel and makes the outputs have a color shift.
\paragraph{\bf Imperceptibility of adversarial perturbations.}
The threat posed by adversarial perturbations partly stems from their invisibility. 
To verify the invisibility of adversarial perturbations generated by our adversarial attacks, 
we conducted a quantitative analysis of the adversarial examples generated 
for the discussed UWIE models.

Table~\ref{tab:imp} demonstrates the PSNR between adversarial examples and their corresponding
original examples with attack parameterized with iteration $T=5$. 
For each examined learning-based UWIE model, the results show that the PSNR between original examples and adversarial examples remains at a considerably high level. Even the most aggressive adversarial attack ($\epsilon = 8/255$),
cannot reduce PSNR to an unacceptable level. Such experimental results strongly demonstrate the invisibility of adversarial perturbations and also showcase the threat posed by adversarial samples.

\section{Conclusions}
In this paper, we first conducted a comprehensive study of the adversarial robustness of learning-based UWIE methods.
On the basis of the great impact of color space on UWIE models, we also design two effective UWIE-oriented adversarial attack methods, Pixel Attack, and Color
Shift Attack, targeting different color spaces. We conducted adversarial attacks on five well-designed learning-based UWIE models. The experiments showed the ubiquity of vulnerabilities of learning-base models to adversarial attacks 
and we attempted to summarize the visual impact of adversarial examples on the output of the UWIE model.
We also adopted an adversarial training method to defend these models and successfully reduced the threat of adversarial examples to learning-based UWIE models.

We shed light on the adversarial vulnerability to learning-based UWIE 
models and proposed the evaluation metrics for the robustness of UWIE models, a new evaluation dimension of UWIE models, which brings numerous interesting new directions. 


 \bibliographystyle{IEEEtran}
 \bibliography{sample-base}

\begin{thebibliography}{10}
\providecommand{\url}[1]{#1}
\csname url@samestyle\endcsname
\providecommand{\newblock}{\relax}
\providecommand{\bibinfo}[2]{#2}
\providecommand{\BIBentrySTDinterwordspacing}{\spaceskip=0pt\relax}
\providecommand{\BIBentryALTinterwordstretchfactor}{4}
\providecommand{\BIBentryALTinterwordspacing}{\spaceskip=\fontdimen2\font plus
\BIBentryALTinterwordstretchfactor\fontdimen3\font minus
  \fontdimen4\font\relax}
\providecommand{\BIBforeignlanguage}[2]{{%
\expandafter\ifx\csname l@#1\endcsname\relax
\typeout{** WARNING: IEEEtran.bst: No hyphenation pattern has been}%
\typeout{** loaded for the language `#1'. Using the pattern for}%
\typeout{** the default language instead.}%
\else
\language=\csname l@#1\endcsname
\fi
#2}}
\providecommand{\BIBdecl}{\relax}
\BIBdecl

\bibitem{mclellan2015sustainability}
B.~C. McLellan, ``Sustainability assessment of deep ocean resources,''
  \emph{Procedia Environmental Sciences}, vol.~28, pp. 502--508, 2015.

\bibitem{alsakar2023underwater}
Y.~M. Alsakar, N.~A. Sakr, S.~El-Sappagh, T.~Abuhmed, and M.~Elmogy,
  ``Underwater image restoration and enhancement: A comprehensive review of
  recent trends, challenges, and applications,'' 2023.

\bibitem{lin2022global}
R.~Lin, J.~Liu, R.~Liu, and X.~Fan, ``Global structure-guided learning
  framework for underwater image enhancement,'' \emph{The Visual Computer},
  vol.~38, no.~12, pp. 4419--4434, 2022.

\bibitem{xue2023investigating}
X.~Xue, Z.~Li, L.~Ma, Q.~Jia, R.~Liu, and X.~Fan, ``Investigating intrinsic
  degradation factors by multi-branch aggregation for real-world underwater
  image enhancement,'' \emph{Pattern Recognition}, vol. 133, p. 109041, 2023.

\bibitem{li2019underwater}
C.~Li, C.~Guo, W.~Ren, R.~Cong, J.~Hou, S.~Kwong, and D.~Tao, ``An underwater
  image enhancement benchmark dataset and beyond,'' \emph{IEEE Transactions on
  Image Processing}, vol.~29, pp. 4376--4389, 2019.

\bibitem{jiang2022target}
Z.~Jiang, Z.~Li, S.~Yang, X.~Fan, and R.~Liu, ``Target oriented perceptual
  adversarial fusion network for underwater image enhancement,'' \emph{IEEE
  Transactions on Circuits and Systems for Video Technology}, vol.~32, no.~10,
  pp. 6584--6598, 2022.

\bibitem{liu2022twin}
R.~Liu, Z.~Jiang, S.~Yang, and X.~Fan, ``Twin adversarial contrastive learning
  for underwater image enhancement and beyond,'' \emph{IEEE Transactions on
  Image Processing}, vol.~31, pp. 4922--4936, 2022.

\bibitem{guo2020zero}
C.~Guo, C.~Li, J.~Guo, C.~C. Loy, J.~Hou, S.~Kwong, and R.~Cong,
  ``Zero-reference deep curve estimation for low-light image enhancement,'' in
  \emph{IEEE/CVF conference on computer vision and pattern recognition}, 2020,
  pp. 1780--1789.

\bibitem{li2021learning}
C.~Li, C.~Guo, and C.~C. Loy, ``Learning to enhance low-light image via
  zero-reference deep curve estimation,'' \emph{IEEE Transactions on Pattern
  Analysis and Machine Intelligence}, vol.~44, no.~8, pp. 4225--4238, 2021.

\bibitem{jiang2022unsupervised}
Q.~Jiang, Y.~Mao, R.~Cong, W.~Ren, C.~Huang, and F.~Shao, ``Unsupervised
  decomposition and correction network for low-light image enhancement,''
  \emph{IEEE Transactions on Intelligent Transportation Systems}, vol.~23,
  no.~10, pp. 19\,440--19\,455, 2022.

\bibitem{ijcai2022p160}
Y.~Liang, E.~Huang, Z.~Zhang, Z.~Su, and D.~Wang, ``Feature dense relevance
  network for single image dehazing,'' in \emph{International Joint Conference
  on Artificial Intelligence}, 7 2022, pp. 1144--1150.

\bibitem{wang2022adaptive}
Z.~Wang, F.~Li, R.~Cong, H.~Bai, and Y.~Zhao, ``Adaptive feature fusion network
  based on boosted attention mechanism for single image dehazing,''
  \emph{Multimedia Tools and Applications}, vol.~81, no.~8, pp.
  11\,325--11\,339, 2022.

\bibitem{zhou2023underwater}
J.~Zhou, Q.~Liu, Q.~Jiang, W.~Ren, K.-M. Lam, and W.~Zhang, ``Underwater
  camera: Improving visual perception via adaptive dark pixel prior and color
  correction,'' \emph{International Journal of Computer Vision}, pp. 1--19,
  2023.

\bibitem{zhou2024hclr}
J.~Zhou, J.~Sun, C.~Li, Q.~Jiang, M.~Zhou, K.-M. Lam, W.~Zhang, and X.~Fu,
  ``Hclr-net: Hybrid contrastive learning regularization with locally
  randomized perturbation for underwater image enhancement,''
  \emph{International Journal of Computer Vision}, pp. 1--25, 2024.

\bibitem{ghani2017automatic}
A.~S.~A. Ghani and N.~A.~M. Isa, ``Automatic system for improving underwater
  image contrast and color through recursive adaptive histogram modification,''
  \emph{Computers and electronics in agriculture}, vol. 141, pp. 181--195,
  2017.

\bibitem{sankpal2019underwater}
S.~Sankpal and S.~Deshpande, ``Underwater image enhancement by rayleigh
  stretching with adaptive scale parameter and energy correction,'' in
  \emph{Computing, Communication and Signal Processing}, 2019, pp. 935--947.

\bibitem{priyadharsini2018wavelet}
T.~Priyadharsini~R, Sree~Sharmila and V.~Rajendran, ``A wavelet transform based
  contrast enhancement method for underwater acoustic images,''
  \emph{Multidimensional Systems and Signal Processing}, vol.~29, pp.
  1845--1859, 2018.

\bibitem{wang2022periodic}
J.~Wang, M.~Wan, G.~Gu, W.~Qian, K.~Ren, Q.~Huang, and Q.~Chen, ``Periodic
  integration-based polarization differential imaging for underwater image
  restoration,'' \emph{Optics and Lasers in Engineering}, vol. 149, p. 106785,
  2022.

\bibitem{yan2022attention}
X.~Yan, W.~Qin, Y.~Wang, G.~Wang, and X.~Fu, ``Attention-guided dynamic
  multi-branch neural network for underwater image enhancement,''
  \emph{Knowledge-Based Systems}, vol. 258, p. 110041, 2022.

\bibitem{islam2020fast}
M.~J. Islam, Y.~Xia, and J.~Sattar, ``Fast underwater image enhancement for
  improved visual perception,'' \emph{IEEE Robotics and Automation Letters},
  vol.~5, no.~2, pp. 3227--3234, 2020.

\bibitem{chen2021underwater}
X.~Chen, P.~Zhang, L.~Quan, C.~Yi, and C.~Lu, ``Underwater image enhancement
  based on deep learning and image formation model,'' 2021.

\bibitem{9747758}
Z.~Fu, X.~Lin, W.~Wang, Y.~Huang, and X.~Ding, ``Underwater image enhancement
  via learning water type desensitized representations,'' in \emph{ICASSP 2022
  - 2022 IEEE International Conference on Acoustics, Speech and Signal
  Processing (ICASSP)}, 2022, pp. 2764--2768.

\bibitem{wei2022uhd}
Y.~Wei, Z.~Zheng, and X.~Jia, ``Uhd underwater image enhancement via
  frequency-spatial domain aware network,'' in \emph{the Asian Conference on
  Computer Vision}, 2022, pp. 299--314.

\bibitem{10155564}
R.~Cong, W.~Yang, W.~Zhang, C.~Li, C.-L. Guo, Q.~Huang, and S.~Kwong, ``Pugan:
  Physical model-guided underwater image enhancement using gan with
  dual-discriminators,'' \emph{IEEE Transactions on Image Processing}, vol.~32,
  pp. 4472--4485, 2023.

\bibitem{goodfellow2014explaining}
I.~J. Goodfellow, J.~Shlens, and C.~Szegedy, ``Explaining and harnessing
  adversarial examples,'' \emph{arXiv preprint arXiv:1412.6572}, 2014.

\bibitem{madry2018towards}
A.~Madry, A.~Makelov, L.~Schmidt, D.~Tsipras, and A.~Vladu, ``Towards deep
  learning models resistant to adversarial attacks,'' in \emph{International
  Conference on Learning Representations}, 2018.

\bibitem{wei2023cfa}
Z.~Wei, Y.~Wang, Y.~Guo, and Y.~Wang, ``Cfa: Class-wise calibrated fair
  adversarial training,'' in \emph{Proceedings of the IEEE/CVF Conference on
  Computer Vision and Pattern Recognition}, 2023, pp. 8193--8201.

\bibitem{croce2021mind}
F.~Croce and M.~Hein, ``Mind the box: $l_1$-apgd for sparse adversarial attacks
  on image classifiers,'' in \emph{International conference on machine
  learning}, 2021.

\bibitem{yu2022towards}
Y.~Yu, W.~Yang, Y.-P. Tan, and A.~C. Kot, ``Towards robust rain removal against
  adversarial attacks: A comprehensive benchmark analysis and beyond,'' in
  \emph{IEEE/CVF Conference on Computer Vision and Pattern Recognition}, 2022,
  pp. 6013--6022.

\bibitem{gui2023adversarial}
J.~Gui, X.~Cong, C.~Peng, Y.~Y. Tang, and J.~T.-Y. Kwok, ``Adversarial attack
  and defense for dehazing networks,'' \emph{arXiv preprint arXiv:2303.17255},
  2023.

\bibitem{zhang2023hierarchical}
D.~Zhang, C.~Wu, J.~Zhou, W.~Zhang, C.~Li, and Z.~Lin, ``Hierarchical attention
  aggregation with multi-resolution feature learning for gan-based underwater
  image enhancement,'' \emph{Engineering Applications of Artificial
  Intelligence}, vol. 125, p. 106743, 2023.

\bibitem{zhuhierarchical}
G.~Zhu, L.~Ma, X.~Fan, and R.~Liu, ``Hierarchical bilevel learning with
  architecture and loss search for hadamard-based image restoration,'' 2022.

\bibitem{sharma2023wavelength}
P.~Sharma, I.~Bisht, and A.~Sur, ``Wavelength-based attributed deep neural
  network for underwater image restoration,'' \emph{ACM Transactions on
  Multimedia Computing, Communications and Applications}, vol.~19, no.~1, pp.
  1--23, 2023.

\bibitem{zhang2023rex}
D.~Zhang, J.~Zhou, W.~Zhang, Z.~Lin, J.~Yao, K.~Polat, F.~Alenezi, and
  A.~Alhudhaif, ``Rex-net: A reflectance-guided underwater image enhancement
  network for extreme scenarios,'' \emph{Expert Systems with Applications}, p.
  120842, 2023.

\bibitem{zhang2023waterflow}
Z.~Zhang, Z.~Jiang, J.~Liu, X.~Fan, and R.~Liu, ``Waterflow: heuristic
  normalizing flow for underwater image enhancement and beyond,'' in \emph{ACM
  International Conference on Multimedia}, 2023, pp. 7314--7323.

\bibitem{zhou2022underwater}
J.~Zhou, T.~Yang, W.~Chu, and W.~Zhang, ``Underwater image restoration via
  backscatter pixel prior and color compensation,'' \emph{Engineering
  Applications of Artificial Intelligence}, vol. 111, p. 104785, 2022.

\bibitem{papernot2017practical}
N.~Papernot, P.~McDaniel, I.~Goodfellow, S.~Jha, Z.~B. Celik, and A.~Swami,
  ``Practical black-box attacks against machine learning,'' in \emph{Asia
  conference on computer and communications security}, 2017, pp. 506--519.

\bibitem{szegedy2013intriguing}
C.~Szegedy, W.~Zaremba, I.~Sutskever, J.~Bruna, D.~Erhan, I.~Goodfellow, and
  R.~Fergus, ``Intriguing properties of neural networks,'' \emph{arXiv preprint
  arXiv:1312.6199}, 2013.

\bibitem{moosavi2016deepfool}
S.-M. Moosavi-Dezfooli, A.~Fawzi, and P.~Frossard, ``Deepfool: a simple and
  accurate method to fool deep neural networks,'' in \emph{IEEE conference on
  computer vision and pattern recognition}, 2016, pp. 2574--2582.

\bibitem{carlini2017towards}
N.~Carlini and D.~Wagner, ``Towards evaluating the robustness of neural
  networks,'' in \emph{IEEE symposium on security and privacy}.\hskip 1em plus
  0.5em minus 0.4em\relax Ieee, 2017, pp. 39--57.

\bibitem{bai2019hilbert}
Y.~Bai, Y.~Feng, Y.~Wang, T.~Dai, S.-T. Xia, and Y.~Jiang, ``Hilbert-based
  generative defense for adversarial examples,'' in \emph{IEEE/CVF
  International Conference on Computer Vision}, 2019, pp. 4784--4793.

\bibitem{das2017keeping}
N.~Das, M.~Shanbhogue, S.-T. Chen, F.~Hohman, L.~Chen, M.~E. Kounavis, and
  D.~H. Chau, ``Keeping the bad guys out: Protecting and vaccinating deep
  learning with jpeg compression,'' \emph{arXiv preprint arXiv:1705.02900},
  2017.

\bibitem{mo2022adversarial}
Y.~Mo, D.~Wu, Y.~Wang, Y.~Guo, and Y.~Wang, ``When adversarial training meets
  vision transformers: Recipes from training to architecture,'' \emph{Advances
  in Neural Information Processing Systems}, vol.~35, pp. 18\,599--18\,611,
  2022.

\bibitem{papernot2016distillation}
N.~Papernot, P.~McDaniel, X.~Wu, S.~Jha, and A.~Swami, ``Distillation as a
  defense to adversarial perturbations against deep neural networks,'' in
  \emph{IEEE symposium on security and privacy}, 2016, pp. 582--597.

\bibitem{mustafa2019image}
A.~Mustafa, S.~H. Khan, M.~Hayat, J.~Shen, and L.~Shao, ``Image
  super-resolution as a defense against adversarial attacks,'' \emph{IEEE
  Transactions on Image Processing}, vol.~29, pp. 1711--1724, 2019.

\bibitem{zhai2020s}
L.~Zhai, F.~Juefei-Xu, Q.~Guo, X.~Xie, L.~Ma, W.~Feng, S.~Qin, and Y.~Liu,
  ``It’s raining cats or dogs? adversarial rain attack on dnn perception,''
  \emph{arXiv preprint arXiv:2009.09205}, vol.~2, 2020.

\bibitem{gao2021advhaze}
R.~Gao, Q.~Guo, F.~Juefei-Xu, H.~Yu, and W.~Feng, ``Advhaze: Adversarial haze
  attack,'' \emph{arXiv preprint arXiv:2104.13673}, 2021.

\bibitem{song2023robust}
Z.~Song, Z.~Zhang, K.~Zhang, W.~Luo, Z.~Fan, W.~Ren, and J.~Lu, ``Robust single
  image reflection removal against adversarial attacks,'' in \emph{IEEE/CVF
  Conference on Computer Vision and Pattern Recognition}, 2023, pp.
  24\,688--24\,698.

\bibitem{zhang2024robust}
D.~Zhang, C.~Wu, J.~Zhou, W.~Zhang, Z.~Lin, K.~Polat, and F.~Alenezi, ``Robust
  underwater image enhancement with cascaded multi-level sub-networks and
  triple attention mechanism,'' \emph{Neural Networks}, vol. 169, pp. 685--697,
  2024.

\bibitem{wang2022semantic}
D.~Wang, L.~Ma, R.~Liu, and X.~Fan, ``Semantic-aware texture-structure feature
  collaboration for underwater image enhancement,'' in \emph{International
  Conference on Robotics and Automation}, 2022, pp. 4592--4598.

\bibitem{zhou2023ugif}
J.~Zhou, B.~Li, D.~Zhang, J.~Yuan, W.~Zhang, Z.~Cai, and J.~Shi, ``Ugif-net: An
  efficient fully guided information flow network for underwater image
  enhancement,'' \emph{IEEE Transactions on Geoscience and Remote Sensing},
  2023.

\bibitem{spl}
M.~S. Sarfraz, C.~Seibold, H.~Khalid, and R.~Stiefelhagen, ``Content and colour
  distillation for learning image translations with the spatial profile loss,''
  in \emph{British Machine Vision Conference}, 2019.

\bibitem{zhang2019theoretically}
H.~Zhang, Y.~Yu, J.~Jiao, E.~Xing, L.~El~Ghaoui, and M.~Jordan, ``Theoretically
  principled trade-off between robustness and accuracy,'' in
  \emph{International conference on machine learning}, 2019, pp. 7472--7482.

\bibitem{cong2024underwater}
X.~Cong, J.~Gui, and J.~Hou, ``Underwater organism color fine-tuning via
  decomposition and guidance,'' in \emph{AAAI Conference on Artificial
  Intelligence}, vol.~38, no.~2, 2024, pp. 1389--1398.

\bibitem{wang2004image}
Z.~Wang, A.~C. Bovik, H.~R. Sheikh, and E.~P. Simoncelli, ``Image quality
  assessment: from error visibility to structural similarity,'' \emph{IEEE
  transactions on image processing}, vol.~13, no.~4, pp. 600--612, 2004.

\end{thebibliography}

\end{document}